\documentclass[letterpaper,11pt]{article}
\usepackage[authoryear]{natbib}
\usepackage{tabularx}
\usepackage{amsmath}
\usepackage{graphicx}
\usepackage[margin=1in,letterpaper]{geometry}
\usepackage[final]{hyperref}
\usepackage[dvipsnames]{xcolor}
\usepackage{amssymb}
\usepackage{booktabs}
\usepackage{longtable}
\usepackage{array}
\usepackage{multirow}
\usepackage{wrapfig}
\usepackage{float}
\usepackage{colortbl}
\usepackage{pdflscape}
\usepackage{tabu}
\usepackage{threeparttable}
\usepackage{threeparttablex}
\usepackage[normalem]{ulem}
\usepackage{makecell}
\usepackage{multirow}
\usepackage{subfig}

\newcommand{\source}[1]{
{{\footnotesize \\ Note: #1}}
}

\hypersetup{
	colorlinks=true,       
	linkcolor=blue,        
	citecolor=blue,        
	filecolor=magenta,     
	urlcolor=blue         
}

\begin{document}

\title{\textbf{Sizing the Risk: Kelly, VIX, and Hybrid Approaches in Put-Writing on Index Options}}

\author{Maciej Wysocki$^{1}$\thanks{Corresponding author: m.wysocki9@uw.edu.pl, 44/50 Dluga, Warsaw, Poland. ORCID: 0000-0002-1693-1438} \\ \\ 
\small $^{1}$Department of Quantitative Finance and Machine Learning \\ 
\small Faculty of Economic Sciences, University of Warsaw \\
\small Quantitative Finance Research Group}
\date{}

\maketitle

\renewcommand*\abstractname{Abstract}
\begin{abstract}

This paper examines systematic put-writing strategies applied to S\&P 500 Index options, with a focus on position sizing as a key determinant of long-term performance. Despite the well-documented volatility risk premium, where implied volatility exceeds realized volatility, the practical implementation of short-dated volatility-selling strategies remains underdeveloped in the literature. This study evaluates three position sizing approaches: the Kelly criterion, VIX-based volatility regime scaling, and a novel hybrid method combining both. Using SPXW options with expirations from 0 to 5 days, the analysis explores a broad design space, including moneyness levels, volatility estimators, and memory horizons. Results show that ultra-short-dated, far out-of-the-money options deliver superior risk-adjusted returns. The hybrid sizing method consistently balances return generation with robust drawdown control, particularly under low-volatility conditions such as those seen in 2024. The study offers new insights into volatility harvesting, introducing a dynamic sizing framework that adapts to shifting market regimes. It also contributes practical guidance for constructing short-dated option strategies that are robust across market environments. These findings have direct applications for institutional investors seeking to enhance portfolio efficiency through systematic exposure to volatility premia.


\end{abstract}

\small{\noindent\textbf{Keywords:} S\&P 500 Index Options, Implied Volatility, Volatility Index, VIX, Short Put\\}


\newpage

\section{Introduction}
The systematic harvesting of volatility risk premia through options strategies has emerged as one of the most compelling areas of modern quantitative finance. At its core, volatility reflects the market's collective assessment of future price uncertainty, an assessment that frequently diverges from subsequently realized movements, creating persistent opportunities for informed investors \citep{bardgett_inferring_2019, eraker_volatility_2021}.

The concept of the volatility risk premium (VRP) is grounded in a well-documented empirical phenomenon: implied volatility, as embedded in option prices, tends to exceed realized volatility across most time horizons and market regimes \citep{christensen_relation_1998}. This persistent discrepancy reflects compensation demanded by market participants for bearing volatility risk and resembles the business model of insurance firms, which routinely collect premiums exceeding expected payouts. In the context of options markets, systematic sellers of volatility, such as put-writing strategies, may be able to capture this premium over time \citep{malkiel_option_2018, patel_cash_secured_2024}.

Put-writing strategies, which involve the systematic sale of put options on broad equity indices, offer one of the most direct mechanisms for extracting the volatility risk premium. By selling a put, the investor effectively provides insurance against downside market movements, earning a premium in exchange for assuming tail risk. The appeal of this strategy lies in the statistical tendency for most options to expire worthless or near-worthless, allowing the seller to retain a large portion of the premium \citep{black_35-year_2022}. However, this edge must be weighed against the risk of infrequent but potentially severe losses during sharp market declines.

As discussed, the theoretical basis for volatility-selling strategies draws upon several strands of financial theory. Yet, the practical implementation of these strategies entails numerous complex design decisions that can materially impact outcomes. For instance, should one prioritize near-the-money options that offer higher premiums but carry greater downside risk, or deeper out-of-the-money options that provide a more favorable risk-reward profile? Similarly, should the strategy focus on ultra-short-dated options, which benefit from rapid time decay, or on slightly longer-dated options with more favorable payoff profiles? These choices are further complicated by questions of capital allocation, volatility forecasting methodology, and responsiveness to evolving market conditions.

Among these decisions, position sizing is particularly consequential. The nonlinear payoff structure of options introduces fat-tailed risk exposures, meaning that improper sizing can result in catastrophic drawdowns despite a strategy’s positive expected return. Conventional portfolio theory often falls short in this context due to its assumptions of normality and linear exposure.

The Kelly criterion \citep{kelly_jr_new_1956}, originally developed in information theory, offers a mathematically optimal approach to position sizing by maximizing long-term capital growth \citep{thorp_kelly_2008}. When adapted to financial markets, it balances return maximization with the risk of bankruptcy. However, effective application of Kelly sizing requires precise estimates of expected return and variance, which are themselves highly sensitive to option characteristics and prevailing market regimes.

In contrast, volatility regime-based sizing seeks to condition exposure on observed market volatility levels, typically using indicators such as the VIX to assess the current risk environment \citep{malkiel_option_2018}. This approach recognizes that volatility-selling strategies tend to perform better when implied volatility is elevated and more accurately compensates for tail risk. During calmer market periods, when the risk-reward profile deteriorates, these models recommend reduced exposure.

This study addresses these gaps by conducting a comprehensive evaluation of put-writing strategies across three position sizing methodologies: (i) Monte Carlo-based implementations of the Kelly criterion; (ii) VIX-based volatility regime sizing; and (iii) a hybrid approach that blends both frameworks. Each sizing methodology is examined across a wide grid of strategy configurations, including time-to-expiration windows from same-day to five days, moneyness levels from at-the-money to 10\% out-of-the-money, and multiple volatility estimation techniques with varying memory horizons.

This research contributes to the literature in several important ways. First, it introduces a novel hybrid sizing methodology that combines Monte Carlo simulation-based forward-looking estimates with real-time volatility regime signals. This technique offers practical advantages for managing tail risk while maintaining attractive growth prospects. Second, it presents a novel portfolio construction scheme with frequent rollovers according to options expiry. This offers very unique insights into the behavior of options approaching expiry, which is available due to the use of SPXW options expiring almost every trading day, that are gaining in popularity and trading volume \citep{beckmeyer2023retail}. Third, it presents the most thorough analysis to date of how multiple design elements — such as option selection, time horizon, and volatility input — interact to influence risk-adjusted performance. Fourth, it offers concrete implementation guidance for investors seeking to incorporate systematic volatility-selling strategies within diversified portfolios.

The findings hold both theoretical and practical relevance. On the theoretical front, the results deepen our understanding of volatility risk premium dynamics and broaden the available scope of ideas for quantitative trading strategies. Practically, they offer actionable insights for portfolio managers navigating increasingly sophisticated and competitive options markets.

The remainder of this paper is organized as follows: Section~2 outlines the methodology; Section~3 describes the backtesting procedures and evaluation metrics; Section~4 presents the data; Section~5 discusses the in-sample results; Section~6 details the out-of-sample performance; and Section~7 concludes.

\subsection{Relevant Literature}

The merits of option-writing strategies have been extensively discussed in the literature. \citet{patel_cash_secured_2024} offer a comprehensive rationale for the attractiveness of put-writing strategies, emphasizing their potential to generate consistent excess returns. The authors attribute the outperformance of the CBOE S\&P 500 PutWrite Index (PUT) to exposure to the VRP, suggesting that systematically harvesting this premium can lead to profitable outcomes. However, they caution that such strategies are not without practical limitations. In particular, trading frictions such as transaction costs and margin requirements can significantly degregade returns, posing real challenges to implementation.

In a related study, \citet{hong_profitability_2018} analyze the impact of trading frictions on the profitability of volatility-based strategies using S\&P 500 Index options. The authors evaluate a variety of put-writing strategies, differentiated by time to expiration - 30 and 60 days-to-expiry (DTE) held to maturity, as well as a 45 DTE strategy closed at 15 DTE — and moneyness, including 10\% OTM puts. Their empirical analysis spans two sub-periods, 2002–2007 and 2008–2013, to capture the contrasting dynamics before and after the global financial crisis. While the strategies exhibited strong performance on a risk-adjusted basis, the study underscores that bid-ask spreads represent a non-negligible source of slippage, potentially causing realized returns to deviate materially from theoretical backtests. Additionally, the authors highlight that the capital tied up in margin requirements is substantial and must be carefully considered by investors pursuing such strategies.

One of the most comprehensive reviews of S\&P 500 Index option-writing strategies is provided by \citet{black_35-year_2022}, who assess the long-term performance of various CBOE option-writing indices. The study presents an accessible evaluation of benchmark indices, including those linked to covered calls and put-writing strategies, alongside a discussion of the embedded VRP in the S\&P 500 Index. The authors find that all examined option-writing indices, including the PUT index, delivered statistically significant positive abnormal returns over the study horizon. These findings support the notion that systematic option selling can be profitable, primarily due to the persistent spread between implied and realized volatility. Nevertheless, the authors emphasize that these strategies entail meaningful risk exposures and are highly sensitive to market timing.

\citet{ungar_cash_secured_2009} similarly evaluate the performance of put-writing strategies based on the CBOE S\&P 500 PutWrite Index. Their results indicate that while these strategies can offer attractive risk-adjusted returns, they tend to underperform in strong bull markets. The study concludes that the efficacy of put-writing is heavily conditioned by prevailing market environments, with superior outcomes observed during periods when implied volatility significantly exceeds realized volatility—conditions in which puts are priced rich.

A complementary perspective is provided by \citet{schwalbach_analysis_2018}, who conduct a granular analysis of fully collateralized put-writing strategies using SPDR S\&P 500 ETF Trust (SPY) options. Their study explores a broad set of strategy configurations across different maturities (ranging from 30 to 365 days) and moneyness levels (from 10\% OTM to 10\% in-the-money (ITM)). The authors report a clear trade-off between return and risk: longer-dated options were associated with lower returns and higher volatility, while strategies closer to at-the-money levels offered higher returns but also greater drawdowns. Overall, the 5\% OTM put-writing strategies were identified as optimal from a risk-return perspective, balancing moderate risk with appealing return potential.

\section{Methodology}
The strategies examined in this paper are based on systematically writing put options, which can be interpreted as selling insurance in exchange for a premium. A short put position represents a directional exposure: if the underlying asset's price falls below the option’s strike price at expiration, the writer is obligated to cover the difference between the strike and the underlying price. Conversely, if the underlying price remains above the strike at expiration, the option expires worthless, and the writer retains the full premium as profit.

While selling puts can generate consistent income, it entails significant risk, as sharp declines in the underlying asset can lead to substantial losses, potentially wiping out the entire portfolio. To address this risk, the paper introduces various position sizing methodologies designed to limit exposure and enhance risk management. Specifically, strategies based on the Kelly criterion \citep{kelly_jr_new_1956, rotando_kelly_1992}, the VIX-rank, and a hybrid approach combining both techniques are proposed and evaluated.

\subsection{Investment Strategies}
This paper introduces a novel portfolio construction methodology that, to the author's knowledge, has not been explored in the existing literature. Leveraging the frequent expirations of SPXW options, it is now possible to adjust and initiate short-dated option positions on a daily basis, maintaining constant exposure to very near-term maturities. This creates an opportunity to systematically exploit theta decay, the well-documented phenomenon whereby an option's time value diminishes at an accelerating rate as expiration approaches, particularly for near-the-money options \citep{tannous_expected_2008}.

To capitalize on this effect, the following portfolio construction algorithm is proposed:

\begin{enumerate} 
    \item Select a target days-to-expiration (DTE) value, denoted as $N$. 
    \item Each trading day, beginning 15 minutes after market open, scan available option maturities to identify the expiration date closest to $N$ days. 
    \begin{enumerate} 
        \item If no position is currently open, initiate a short put position using options with DTE closest to $N$. 
        \item If a position is already open, compare the current position's DTE to the DTE of newly available options: 
            \begin{itemize} 
            \item If the existing position's DTE remains closer to $N$, maintain the position. \item Otherwise, close the current position and open a new short position in options closer to $N$. 
            \end{itemize} 
    \end{enumerate} 
\end{enumerate}

Depending on the selected target DTE ($N$), the strategy may either roll positions frequently without holding to expiration, or, for low $N$ values (such as 1 or 0 DTE), hold positions until maturity and reopen them the next day. Thus, the methodology dynamically adjusts between active rolling and passive holding, depending on the proximity of available maturities to the target $N$.

In addition to the portfolio construction algorithm, each strategy adheres to a consistent set of rules governing contract selection:

\begin{itemize} 
    \item \textbf{Position sizing}: The number of contracts sold is determined according to one of three methodologies: the Kelly criterion, the VIX-rank approach, or a hybrid of the two. \item \textbf{Strike selection}: The strike closest to the target moneyness level is selected, subject to the constraint that only strikes divisible by 25 are eligible. 
    \item \textbf{Rollover timing}: Position adjustments are permitted no earlier than 15 minutes after market open and no later than 30 minutes before market close. 
\end{itemize}

The restriction to strikes divisible by 25 serves to enhance the realism of the strategy, reflecting the fact that although finer strikes are available, these contracts are often less liquid and could introduce challenges in performance evaluation.

For the purpose of strike selection, moneyness is defined as follows:

\begin{equation} 
M_{put} = K e^{-r \tau} - S, 
\label{eq:put-moneyness} 
\end{equation}
where $S$ denotes the spot price of the underlying asset, $K$ represents the strike price, $r$ is the risk-free interest rate, and $\tau$ is the time to expiration expressed in years. This study examines four discrete moneyness levels: $\{0\%, 2\%, 5\%, 10\%\}$, corresponding to positions ranging from at-the-money to deep out-of-the-money.

\subsection{Kelly Criterion Sizing}\label{section:kelly}
The Kelly criterion \citep{kelly_jr_new_1956} provides an optimal bet sizing strategy that maximizes the growth rate of wealth. Originally formulated within the framework of information theory, it has been extensively adopted in gambling, portfolio management, and algorithmic trading contexts.

Consider a sequence of independent binary outcomes (wins or losses) associated with a repeated investment opportunity. The objective is to maximize the expected logarithm of final wealth, corresponding to maximizing the long-term geometric mean return. The expected log-growth function $G(f)$, given a bet fraction $f$, is:

\begin{equation}
    G(f) = p \cdot \ln(1+bf) + (1-p) \cdot \ln(1-f),
    \label{eq:kelly-growth}
\end{equation}
where $p$ denotes the probability of a win, $b$ is the net fractional payoff on a successful bet, and $f$ is the fraction of current capital wagered. The model assumes that in the event of a loss, the wagered amount is entirely forfeited. Taking a derivative of $G(f)$ with respect to $f$ and setting it to zero yields the optimal fraction:

\begin{equation}
    f^*=\frac{bp-(1-p)}{b}=\frac{p(b+1)-1}{b}
\end{equation}
which is positive if and only if the expected return is positive, ensuring rational investment behavior. A rigorous proof and a more detailed explanation can be found in Kelly's original paper \citep{kelly_jr_new_1956}. 

In financial applications, where partial losses, the classical Kelly formula can be extended. Following the generalizations proposed by \citet{rotando_kelly_1992} and \citet{thorp_kelly_2008}, the optimal Kelly fraction is:

\begin{equation}
    f^* = \frac{p}{a} - \frac{1-p}{b},
    \label{eq:kelly-investment-fraction}
\end{equation}
where $p$ is the probability of a win, $a$ is the odds of a negative outcome and $b$ is the odds of a positive outcome. This formula considers partial losses, thus is more suitable for investment purposes. 

While the Kelly criterion is mathematically optimal under the assumption of infinite repetitions and known probabilities, several practical considerations must be accounted for:

\begin{itemize} 
    \item The accuracy of the input parameters is critical, and wrong assumptions can lead to suboptimal or even catastrophic outcomes. 
    \item The Kelly strategy maximizes expected logarithmic growth but does not incorporate risk aversion. Consequently, it often leads to highly leveraged, high-volatility portfolio trajectories in the short run \citep{thorp_understanding_2011, ziemba_understanding_2016}. 
    \item To mitigate the risk of significant drawdowns and improve practical robustness, practitioners often employ fractional Kelly strategies, wagering a fixed proportion (e.g., half-Kelly) of the theoretically optimal amount. As shown by \citet{maclean_long-term_2010}, fractional Kelly strategies reduce volatility more than they proportionally reduce expected growth, thus offering a pragmatic trade-off between growth and risk control. 
\end{itemize}

Accordingly, in this study, the Kelly sizing is applied with considerations for both model estimation risk and capital preservation, ensuring that the strategy remains viable under realistic market conditions.

The practical application of the Kelly criterion in equation \eqref{eq:kelly-investment-fraction} requires three key inputs: the probability of a positive outcome $p$, the fractional gain conditional on a favorable outcome $b$, and the fractional loss conditional on an unfavorable outcome $a$. To estimate these parameters, a Monte Carlo simulation framework is employed under the assumption that the underlying asset price $S_t$ evolves according to a geometric Brownian motion (GBM). Specifically, the stochastic differential equation governing $S_t$ is:
\begin{equation} 
	\frac{dS_t}{S_t} = \mu dt + \sigma dW_t, 
\end{equation}
here $W_t$ is a standard Wiener process, $\mu$ denotes the constant drift rate, and $\sigma$ denotes the constant volatility. The closed-form solution to this stochastic differential equation, given initial condition $S_0$, is:

\begin{equation}
	S_t = S_0 \exp\left( \left( \mu - \frac{\sigma^2}{2} \right)t + \sigma W_t \right). 
\end{equation}

For practical simulation, a discrete-time approximation over a small time step $\Delta t$ (assuming 252 trading days per year) is used. The dynamics of the discretized log-returns are modeled as:

\begin{equation} 
	\log\left(\frac{S_{t+\Delta t}}{S_t}\right) = \left( r - q - \frac{\sigma^2}{2} \right)\Delta t + \sigma \sqrt{\Delta t} Z, 
\end{equation}
where $r$ is the risk-free interest rate, $q$ is the continuous dividend yield, $\sigma$ is volatility of the asset and $Z \sim \mathcal{N}(0,1)$ is a standard normal random variable. Consequently, the simulated asset price after $n$ discrete steps is:

\begin{equation} 
	S_{t+n\Delta t} = S_0 \exp\left( \sum_{i=1}^n \left( \left( r - q - \frac{\sigma^2}{2} \right)\Delta t + \sigma \sqrt{\Delta t} Z_i \right) \right), 
\end{equation} 
where ${Z_i}_{i=1}^n$ are independent and identically distributed standard normal draws.

The terminal asset prices $S_T$ at maturity are used to compute the final valuation $V_T$ of the short put position, utilizing the Black-Scholes-Merton pricing framework \citep{black_pricing_1973, merton_theory_1973}. For each simulated path, the return from selling the option is computed as:
\begin{equation} 
r_i = \frac{P - V_T^{(i)}}{P}, 
\end{equation}  where $P$ denotes the premium received at initiation ($t=0$) and $V_T^{(i)}$ is the terminal option value along path $i$. The empirical win probability $p$ is estimated by:
\begin{equation} 
p = \frac{1}{N} \sum_{i=1}^N \mathbf{1}{{ r_i > 0 }}, 
\end{equation} 
where $\mathbf{1}{{ \cdot }}$ denotes the indicator function and $N$ is the total number of simulated paths. Similarly, conditional expectations of gains and losses are estimated by:
\begin{align} 
	b &= \mathbb{E}[r_i \mid r_i > 0] \\
	a &= \mathbb{E}[r_i \mid r_i \leq 0]
\end{align}

To account for margin requirements and ensure sufficient collateralization, the margin requirement for a short put position is modeled following brokerage standards\footnote{Margin model based on Interactive Brokers requirements.}:
\begin{equation}
    M(P_t,S_t,K) = P_t + \max(0.15 \cdot S_t - \max(0, K - S_t), 0.1 \cdot S_t)) \cdot 100,
    \label{eq:put-margin}
\end{equation}
where $S_t$ is the underlying asset price at time $t$, $K$ is the strike price, and $P_t$ is the premium received for selling the option.

Given the estimated Kelly fraction $f^*(p,a,b)$ and the margin requirement $M(P_t,S_t,K)$, the position size in contracts $Q_t$ at time $t$ is determined by:

\begin{equation} 
    Q_t = \left\lfloor \frac{PV_t}{M(P_t, S_t, K)} \cdot f^*(p,a,b) \right\rfloor, 
\end{equation} 
where $PV_t$ is the current portfolio value and $\lfloor \cdot \rfloor$ denotes the floor function, ensuring an integer number of contracts.

\subsubsection{Volatility Estimators}\label{section:vol-est}
A key input in the Monte Carlo simulations required for the Kelly sizing methodology described in Section~\ref{section:kelly} is the volatility of the underlying asset's price process. Volatility is the only parameter in the simulation framework that must be externally estimated, and various methods exist to produce this input. This study focuses on three commonly used historical volatility estimators in the context of options trading: the close-to-close estimator, the Garman-Klass estimator, and the Yang-Zhang estimator.

The close-to-close volatility estimator is widely used due to its simplicity, relying solely on historical closing prices. Its statistical properties are well understood, including sampling behavior and bias correction techniques. However, this estimator is relatively inefficient, as it ignores intraday price information and fails to account for known effects such as overnight price jumps. The close-to-close volatility is computed using the following formula:
\begin{equation}
    \sigma^2_C = \frac{1}{T-1}\sum_{t=1}^{T}\left[\log\left(\frac{C_t}{C_{t-1}}\right)-\frac{1}{T}\sum_{t=1}^{T}\log\left( \frac{C_t}{C_{t-1}} \right) \right]^2,
    \label{eq:c2c-vol}
\end{equation}
where $T$ is the number of observations, and $C_t$ denotes the closing price at time $t$.

The Garman-Klass estimator, introduced by \citet{garman_estimation_1980}, incorporates additional price information, specifically, daily high, low, open, and close prices, to improve estimation efficiency over the close-to-close method. The gain in efficiency depends on the sample size and volatility dynamics. The estimator is calculated as:
\begin{equation}
\sigma^2_{GK} = \frac{1}{T} \sum_{t=1}^{T} \left[ \frac{1}{2} \left( \log\left( \frac{H_t}{L_t} \right) \right)^2 - (2\log2 - 1) \left( \log\left( \frac{C_t}{O_t} \right) \right)^2 \right],
\label{eq:garman-klass-vol}
\end{equation}
where $H_t$, $L_t$, $C_t$, and $O_t$ are the high, low, close, and open prices at time $t$, respectively.

The Yang-Zhang estimator, proposed by \cite{yang_driftindependent_2000}, combines multiple components to produce a drift-independent estimator behaving consistently in the presence of opening jumps. It integrates overnight returns (close-to-open, $\sigma^2_O$), intraday volatility (measured by the Rogers-Satchell estimator \citep[$\sigma^2_{RS}$,][]{rogers_estimating_1991}), and the close-to-close volatility estimator ($\sigma^2_C$). It is particularly suitable when opening price jumps are not dominant. The Yang-Zhang estimator is given by:
\begin{equation}
\sigma^2_{YZ} = \sigma_O^2 + k \cdot \sigma^2_C + (1-k) \cdot \sigma^2_{RS},
\label{eq:yz-vol}
\end{equation}
where the components are defined as:
\begin{align}
    \sigma^2_O &= \frac{1}{T-1}\sum_{t=1}^{T}\left[\log\left(\frac{O_t}{C_{t-1}}\right)-\frac{1}{T}\sum_{t=1}^{T}\log\left( \frac{O_t}{C_{t-1}} \right) \right]^2 \\
    \sigma^2_{RS} &= \frac{1}{T}\sum_{t=1}^{T}\left[\log\left(\frac{H_t}{C_{t}}\right)\cdot \log\left(\frac{H_t}{O_{t}}\right) + \log\left(\frac{L_t}{C_{t}}\right) \cdot \log\left(\frac{L_t}{O_{t}}\right) \right] \\
    k &= \frac{0.34}{1.34 + \frac{T + 1}{T - 1}}
\end{align}

All volatility estimates are computed in variance terms. Before being used in the Monte Carlo simulations, they are annualized and transformed into standard deviations by taking the square root. Appropriate scaling factors are applied to ensure consistency across estimators and alignment with the annualized return framework used in the simulations.

\subsection{VIX-Rank Sizing}
The CBOE Volatility Index \citep[VIX,][]{CBOEVIXMethodology2024, CBOEVIXMathMethodology2025} is a forward-looking measure of expected volatility derived from S\&P 500 Index options. The most widely recognized variant reflects market expectations over the next 30 days (VIX), though shorter-term indices, such as the 9-day (VIX9D) and 1-day (VIX1D) volatility indices, are also available. The VIX is considered forward-looking because it is based on implied volatilities rather than historical returns, providing an annualized estimate of the expected magnitude of S\&P 500 Index changes over the selected horizon.

The VIX is often interpreted as a barometer of market sentiment, reflecting perceived risk and investor uncertainty about future conditions. Empirically, the VIX exhibits an inverse relationship with the S\&P 500 Index, where sharp declines in the index are typically accompanied by significant increases in the VIX, and vice versa. Notably, the VIX reacts more strongly to negative index movements than to positive ones, exhibiting asymmetric sensitivity to downside risk \citep{whaley_understanding_2009}.

Given its role as a real-time measure of perceived risk, the VIX can be incorporated into position sizing frameworks to dynamically adjust risk exposure. Specifically, during periods of elevated implied volatility, position sizes should be scaled down to mitigate potential losses, while during periods of low volatility, larger positions may be justified.

The number of options sold at time $t$, denoted $Q_t$, is defined as:

\begin{equation}
    Q_t = \Bigg\lfloor{\frac{PV_t}{M(P_t,S_t,K)} \cdot \left(1 - P_{rank}(VIX_t, W)\right)}\Bigg\rfloor, 
    \label{eq:vix-sizing} 
\end{equation}
where $PV_t$ is the total portfolio value at time $t$, $M(P_t,S_t,K)$ is the per-contract margin requirement as defined in equation \eqref{eq:put-margin}, $VIX_t$ is the current VIX value, and $P_{rank}(VIX_t, W)$ denotes the percentile rank of $VIX_t$ over a lookback window of size $W$.

The percentile rank function is defined as:

\begin{equation} 
    P_{rank}(x, W) = \frac{100}{W \cdot k(x)} \sum_{i=1}^{k(x)} r_i, 
    \label{eq:percentile-rank}
\end{equation}
where $k(x)$ is the frequency of value $x$ within the sample of $W$ observations, and $r_i$ denotes the rank position of each occurrence of $x$ in the sorted array.

Under this framework, a higher VIX percentile corresponds to smaller position sizes, reflecting heightened perceived risk, whereas a lower VIX percentile permits more aggressive allocation. Incorporating the margin requirement ensures that sufficient capital is reserved to satisfy regulatory and broker-imposed collateral constraints. Rounding down to the nearest integer introduces an additional layer of conservatism into the sizing process, although alternative rounding conventions could be considered depending on specific risk management preferences.

\subsection{Kelly Criterion and VIX-Rank Combination Sizing}
A dynamic extension of the Kelly criterion, incorporating volatility information via the VIX-Rank, is proposed to address the inherent path volatility associated with full Kelly betting. Following the framework of fractional Kelly strategies \citep{maclean_long-term_2010}, which aim to retain the favorable growth-optimal properties while reducing equity curve fluctuations, the proposed method dynamically adjusts the investment fraction in response to prevailing market volatility conditions. Rather than applying a static fraction of the Kelly-optimal bet size, the approach utilizes the VIX-Rank, a normalized measure of implied volatility, as a dynamic scaling factor.

Under this framework, during periods of low implied volatility (low VIX-Rank), a larger fraction of the Kelly-optimal position is allocated, taking advantage of more stable market conditions. Conversely, during periods of elevated volatility (high VIX-Rank), the position size is conservatively reduced to preserve capital and mitigate drawdown risks.

Formally, combining the components from equations~\eqref{eq:kelly-investment-fraction} and~\eqref{eq:vix-sizing}, the position size at time $t$ is given by:

\begin{equation} 
    Q_t = \left\lfloor \frac{PV_t}{M(P_t, S_t, K)} \cdot f^*(p, a, b) \cdot \left(1 - P_{\text{rank}}(VIX_t, W)\right) \right\rfloor, 
    \label{eq:kelly-vix-sizing} 
\end{equation} 
where $PV_t$ denotes the portfolio value, $M(P_t, S_t, K)$ is the per-contract margin requirement, $f^*(p,a,b)$ is the optimal Kelly fraction, and $P_{\text{rank}}(VIX_t, W)$ is the normalized VIX-Rank computed over a historical window of size $W$. The floor function ensures an integer number of contracts is traded.

This adaptive sizing mechanism offers several desirable properties. First, it systematically reduces exposure when market uncertainty, as captured by the VIX, is high - an environment typically associated with larger price swings and elevated risk for option writers. Second, by scaling up during periods of calm, the strategy retains the potential for attractive growth in favorable conditions without permanently sacrificing aggressiveness. Finally, because the adjustment is continuous and smooth rather than binary, the resulting equity curve exhibits reduced volatility and improved risk-adjusted performance metrics. By linking position sizing directly to an observable market-based volatility index, the strategy responds flexibly to changing risk environments without requiring explicit market regime classification.

\section{Backtesting and Evaluation}

\subsection{Backtesting Assumptions}
As with any backtesting study, several assumptions had to be made, most of which aim to provide realistic and reliable results. As this study relies on trading and financial data, the relevant commissions had to be taken into consideration in the process of assessing results. All backtests assumed the fee structure reflecting the one of the Interacrtive Brokers brokerage\footnote{See: \href{https://www.interactivebrokers.com/en/pricing/commissions-home.php}{Interactive Brokers Commissions} for more information.}, which is an established, globally-operating firm. It also provides a clear fee structure along with all margin requirements relevant for selling options\footnote{See: \href{https://www.interactivebrokers.com/en/trading/margin-requirements.php}{Interactive Brokers Margin Requirements} for more information.}. Furthermore, the execution costs assume that 50\% of the bid-ask spread is crossed on entry, accounting for the strategy covering half of the bid-ask spread. This is generally in line with how markets operate and reflects the reality of trading options when market-makers are on the other side of the trade. At the beginning of each backtest, each strategy starts with \$5M available capital, which is available for trading and all related costs.

Two main benchmarks were selected for the strategies presented in this research. First, the S\&P 500 Index buy-and-hold (B\&H) strategy emulating a passive investment approach, where an investor buys the index-tracing ETF shares at the beginning of a backtesting period and holds until the end. This strategy demonstrates an indifferent approach, that does not incur substantial trading costs and does not require any maintanance or position management, and is commonly employed by individual investors not interested in trading \citep{COLES2022665}. 

The second benchmark used in this study is the CBOE S\&P 500 PutWrite Index \citep[PUT,][]{cboe_sp_500_putwrite_indices_methodology}, which is a which measures the performance of a passive investment strategy, the "Cboe S\&P 500 Collateralized Put Strategy". This strategy involves selling at-the-money S\&P 500 short put options while maintaining a collateralized position in a Treasury bills account, consisting of one-month and three-month Treasury bills. Put options are sold on a monthly basis. The index serves as a well-established benchmark for put-writing strategies, offering greater specificity than the Buy-and-Hold (B\&H) benchmark.

\subsection{Investment Strategies Evaluation}
Assessing the performance of an investment strategy requires evaluating both its profitability and risk, which encompass returns (or losses) and their variability over time. Hence, the primary focus is on risk-adjusted measures. Each metric utilized in this research was computed individually for each trading plan. These measures are considered industry standards \citep[see e.g.,][]{CFA2020GIPS}, with detailed descriptions available in the literature \citep[Chapter~14]{de2018chapter14}. Specifically, the following definitions were used, assuming 252 trading days per year:

\begin{itemize}
    \item Daily returns represent the percentage change in the value of an investment strategy from one day to the next. Let the asset price on day $i$ of the backtest  be denoted as $p_i$, then:
\begin{equation}
r_i:=\frac{p_i - p_{i-1}}{p_i}
\label{eq:dailyret}
\end{equation}
    \item The Annualized Compounded Rate of Return quantifies the average yearly return of an investment over the backtest period, assuming that profits are reinvested. Let the total number of days in the backtesting period be denoted as $n$, then:
\begin{equation}
aRC:=\prod_{i=1}^{n}(1+r_i^{\frac{252}{n}}) - 1
\label{eq:arc}
\end{equation}
    \item The Annualized Standard Deviation of an investment strategy measures the volatility of returns over a year, providing an estimate of the strategy's risk. It quantifies the potential fluctuation of returns from the expected annual return. Let the average of daily returns be denoted as $\overline{r}$: 
\begin{equation}
aSD:=\sqrt{252} \cdot \sqrt{\frac{1}{n-1}\sum_{i=1}^{n}(r_i - \overline{r})^2}
\label{eq:asd}
\end{equation}
    \item The Maximum Drawdown quantifies the largest observed loss from a peak to a trough before a new peak is reached. It is calculated as the greatest percentage decline in the equity line's value from its highest historical level. Let the total asset value at days $x$ and $y$ be denoted as $v_x$ and $v_y$, respectively, where day $x$ precedes day $y$:
\begin{equation}
MD:=\sup_{(x, y)\in\big\{(t_i, t_j) \in \mathbb{R}: t_i < t_j\big\}}\frac{v_x - v_y}{v_x}
\label{eq:md}
\end{equation}
    \item The Maximum Loss Duration represents the time an investment remains below its previous peak before recovering to a new high, measured in days relative to a full trading year. It is the period between the onset of a maximum drawdown and the recovery to the prior peak. Let $x, y$ as defined in formula \eqref{eq:md}, indicate the days of consecutive local maxima of the total asset value:
\begin{equation}
MLD:=\frac{y-x}{252}
\label{eq:mld}
\end{equation}
    \item The Information Ratio assesses the risk-adjusted performance of an investment strategy. A higher IR indicates that the strategy is generating more excess return per unit of risk, making it more attractive. This measure is considered the most important for evaluating risk-adjusted performance.
\begin{equation}
IR:=\frac{aRC}{aSD}
\label{eq:ir1}
\end{equation}
    \item Value-at-Risk (VaR) measures the potential loss in value of an asset or portfolio over a specified time horizon, at a given confidence level. In this case, the historical VaR of portfolio returns at a 95\% confidence level was used, which corresponds to the 5th percentile of past portfolio returns employed to estimate potential future losses:
\begin{equation}
VaR_{\alpha}(X) := -\inf\big\{x \in \mathbb{R}: F_{X}(x) \geq \alpha \big\}
\label{eq:var}
\end{equation}
    \item Conditional Value-at-Risk measures the expected loss given that the loss exceeds the Value-at-Risk at a specified 95\% confidence level. It provides a more comprehensive risk measure by calculating the average of the worst-case historical strategy losses. CVaR (also referred to as expected shortfall, ES) was estimated as the average of returns below $VaR_{\alpha}(X)$:
\begin{equation}
CVaR_{\alpha}{X}:=-\frac{1}{\alpha}\int_{0}^{\alpha} VaR(\gamma)d\gamma
\label{eq:cvar}
\end{equation}
\end{itemize}

When evaluating the performance of trading strategies, in addition to measuring economic relevance (i.e., profits), we may also be interested in assessing the statistical significance of the results. To accurately examine this, following \cite{bailey_sharpe_2012}, the Probabilistic Sharpe Ratio (PSR) was used. PSR is a reliable methodology for measuring the statistical significance of the Sharpe ratio. Specifically, PSR reflects the probability that the Sharpe ratio ($SR$) of a population of returns exceeds a specified benchmark Sharpe ratio ($SR^*$):
\begin{equation}
    \mathbb{P}\left[SR > SR{^*}\right] = \mathbb{P}\left[\widehat{SR} - Z_{\alpha/2} \hat{\sigma} < SR < \widehat{SR} + Z_{\alpha/2} \widehat{\sigma}\right]
\end{equation}
where $\widehat{SR}$ is the observed Sharpe ratio, $Z_{\alpha/2}$ is the critical value from the standard normal distribution corresponding to the desired significance level $\alpha$, and $\widehat{\sigma}$ represents the standard deviation of the estimated Sharpe ratio. The PSR formula can be then formulated as:
\begin{equation}
    \widehat{PSR}(SR^{*}) = Z\left[ \frac{(\widehat{SR}-SR^{*}) \sqrt{n-1}}{\sqrt{1 - \hat{\gamma}_3 \widehat{SR} + \frac{\hat{\gamma}_4 - 1}{4} \widehat{SR^2}}} \right]
\end{equation}
where $n$ is the number of observed returns in the backtesting period, $\hat{\gamma}_3$ is the skewness of those returns, and $\hat{\gamma}_4$ is the kurtosis of those returns. Rhe PSR provides a more accurate measure of performance by reflecting the likelihood that an observed Sharpe ratio is indicative of the strategy's ability to generate returns rather than being driven by random fluctuations.

\section{Data}
The dataset employed in this study primarily consists of S\&P 500 Index options and their underlying asset, the S\&P 500 Index. All backtests were conducted using 1-minute intraday data, including open, high, low, close (OHLC) prices, as well as bid and ask quotes for both the index and the corresponding options. Additionally, some strategies incorporated data on the VIX and VIX9D indices \citep{CBOEVIXMethodology2024, CBOEVIXMathMethodology2025, CBOESPXVolatilityTermIndices2024}. All data were sourced directly from the Chicago Board Options Exchange (CBOE).

The specific instruments traded in the strategies were SPXW options\footnote{For specifications, see \href{https://www.cboe.com/tradable_products/sp_500/spx_weekly_options/specifications/}{CBOE SPXW website}}, which are weekly options (\textit{weeklies}) listed on the S\&P 500 Index. These options are PM-settled and remain tradeable until 3:00 PM CT on the expiration day. Each contract carries a multiplier of \$100, meaning the notional value equals the index level multiplied by \$100. Due to their PM settlement and increasing trading volume \citep{almeida_0dte_2024}, SPXW options offer high liquidity and frequent trading opportunities, making them particularly well-suited for option-writing strategies.

The risk-free rate used throughout the analysis was the U.S. 3-Month Treasury Bill rate, as published by the Federal Reserve Economic Data (FRED)\footnote{See: \href{https://fred.stlouisfed.org/}{FRED, Federal Reserve Economic Data}}. Data for the PUT Index were obtained directly from the \href{https://www.cboe.com/us/indices/dashboard/put/}{CBOE website}.

As several of the presented strategies involve parameter selection and optimization, the dataset was divided into two subsets: in-sample and out-of-sample. The in-sample data was used to evaluate and compare strategy performance across various parameter configurations. The best-performing parameter sets identified in-sample were then applied to the out-of-sample data to assess their robustness and detect potential overfitting. This approach also facilitates the evaluation of model misspecification and strategy behavior over time. Underperformance in the out-of-sample period may suggest, among other factors, that underlying assumptions were invalid or that the previously exploitable market inefficiencies have diminished. The in-sample period spans from the beginning of 2018 to the end of 2024, while the out-of-sample period covers the full year of 2024.

\begin{table}
\caption{Summary statistics of daily returns of the S\&P500 Index, VIX9D and VIX.}
\label{tab:summary-stats}
\resizebox{\columnwidth}{!}{%
\begin{tabular}{@{}llccccccccccccc@{}}
\toprule
Period & Series & Mean & St.D. & Var & Min & P10 & P25 & P50 & P75 & P90 & Max & Skew & Kurtosis & S-W Test \\ \midrule
\multirow{3}{*}{In-Sample}     & S\&P 500 Index & 0.00046 & 0.01298 & 0.00017 & -0.1202 & -0.0128 & -0.0049 & 0.0009 & 0.0069 & 0.0135 & 0.0942 & -0.509 & 12.878   & 0.881*** \\
                               & VIX9D          & 0.00932 & 0.14923 & 0.02227 & -0.3350 & -0.1301 & -0.0744 & -0.0114 & 0.0642 & 0.1602 & 2.1463 & 3.47   & 33.61    & 0.811*** \\
                               & VIX            & 0.00357 & 0.08749 & 0.00765 & -0.2337 & -0.0787 & -0.0455 & -0.0081 & 0.0355 & 0.0936 & 1.156  & 2.89   & 24.78    & 0.839*** \\ \midrule
\multirow{3}{*}{Out-of-Sample} & S\&P 500 Index & 0.00089 & 0.00797 & 0.00006 & -0.021  & -0.0088 & -0.0029 & 0.001   & 0.0057 & 0.0104 & 0.025  & -0.523 & 1.72     & 0.971*** \\
                               & VIX9D          & 0.01067 & 0.16192 & 0.02622 & -0.3583 & -0.1177 & -0.0611 & -0.0077 & 0.0774 & 0.1445 & 1.6453 & 4.70   & 42.71    & 0.694*** \\
                               & VIX            & 0.00493 & 0.09451 & 0.00893 & -0.2816 & -0.0723 & -0.0352 & -0.0065 & 0.0314 & 0.0837 & 0.7404 & 3.32   & 22.83    & 0.743*** \\ \bottomrule
\end{tabular}%
}
\source{The in-sample dataset comprises observations from January 1, 2018, to December 31, 2023, while the out-of-sample dataset covers the period from January 1, 2024, to December 31, 2024. Returns were computed based on daily closing prices. P10, P25, P50, P75, P90 stand for 10th, 25th, 50th, 75th and 90th percentiles respectively. The S-W Test column contains test statistics for the Shapiro-Wilk normality test. *, **, *** indicate statistical significance on 0.1, 0.05, and 0.01 significance levels respectively.}
\end{table}

Table \ref{tab:summary-stats} presents the summary statistics of daily returns for the S\&P 500 Index, VIX9D, and VIX indices across both the in-sample (2018–2023) and out-of-sample (2024) periods. As expected, the volatility indices exhibit substantially higher mean returns and standard deviations than the underlying S\&P 500 Index, reflecting their reactive nature to market uncertainty. In-sample, VIX9D displays the highest volatility (standard deviation of 0.1492), followed by VIX (0.0875). The returns of the S\&P 500 Index show relatively lower variability. Across both periods, the volatility indices are characterized by pronounced positive skewness and elevated kurtosis, indicating asymmetric return distributions with heavy tails—consistent with their behavior during market stress. The S\&P 500 Index returns exhibit mild negative skewness in both samples and substantially lower kurtosis. The Shapiro-Wilk test rejects normality for all return series at conventional significance levels, further supporting the presence of non-Gaussian features in the data. Notably, VIX9D remains the most volatile and non-normal series, with particularly high kurtosis (42.71) in the out-of-sample period, underlining its responsiveness to short-term market fluctuations.

\begin{figure}[!ht]
\caption{\label{fig:indexvalue} S\&P 500 Index, VIX9D and VIX quotes from 02-01-2018 to 31-12-2024.}
\begin{center}
\includegraphics[width=\columnwidth]{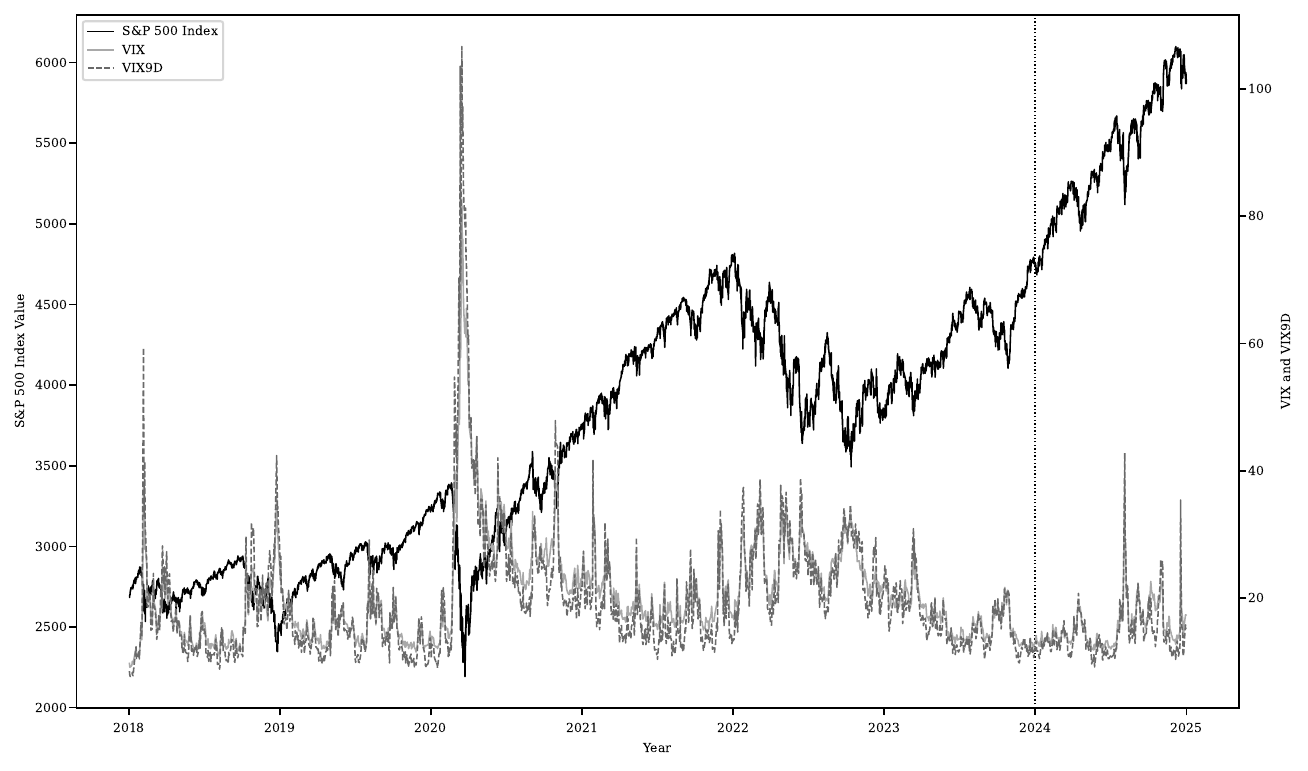}
\end{center}
\footnotesize{Note: The graph presents the average 1-minute bid-ask quotes for the S\&P 500 Index, alongside daily VIX and VIX9D data from the Chicago Board Options Exchange (CBOE). The dashed horizontal line marks the beginning of the out-of-sample period.}
\end{figure}

Figure \ref{fig:indexvalue} displays the values of the S\&P 500 Index alongside the VIX and VIX9D. The visual inspection, supported by the summary statistics in Table \ref{tab:summary-stats}, confirms several well-known stylized facts \citep{cont_empirical_2001, engle_what_2007}. Index returns exhibit heavy tails and gain-loss asymmetry. Both volatility indices display volatility clustering, where high-volatility periods are followed by low-volatility phases. Additionally, volatility mean-reversion is evident, with volatility gradually returning to lower long-term levels after a jump. Comparing the index with its volatility measures reveals the leverage effect, where volatility is negatively correlated with asset returns. Notably, upward return jumps correspond to moderate volatility spikes, while market crashes are associated with sharp volatility surges. This is why the VIX is often referred to as the "fear index", as its elevated values occur during market crashes and periods of economic turmoil.

\section{Results}
The results of this study are primarily presented using heatmaps of information ratios and annualized returns, enabling efficient comparison of multiple strategies across several dimensions. This section focuses on the in-sample performance of all proposed strategies and aims to identify the most promising configurations for subsequent out-of-sample evaluation.

Each strategy is evaluated across a grid of four time-to-expiry (DTE) settings — 0, 1, 3, and 5 days — and four levels of moneyness, defined by the percentage out-of-the-money (OTM): 0\%, 2\%, 5\%, and 10\%. Due to space constraints, detailed tables of in-sample performance are provided in the Appendix. For strategies involving the joint Kelly–VIX sizing methodology, the number of parameter combinations is particularly large, therefore, only a representative subset is included in the Appendix. The complete results, including additional plots, are available in the online Appendix\footnote{Online appendix available at: \url{github.com/Maciej-13/spxw-options-sizing-paper-appendix}}.

\subsection{Strategies with Kelly Criterion Sizing}
The first group of strategies applies the Kelly criterion for position sizing, implemented via Monte Carlo simulations. As described in Section~\ref{section:vol-est}, the key variables are the volatility estimator and the historical lookback window. Three estimators are considered: close-to-close (HV), Garman–Klass (GK), and Yang–Zhang (YZ). Each is evaluated using five lookback windows: 3, 5, 10, 21, and 63 trading days, representing time horizons from very short-term to approximately one quarter.

\begin{figure}[!ht]
\caption{\label{fig:kelly-ir} Information Ratios of Short Put Strategies with Kelly Criterion Sizing}
\begin{center}
\includegraphics[width=0.75\columnwidth]{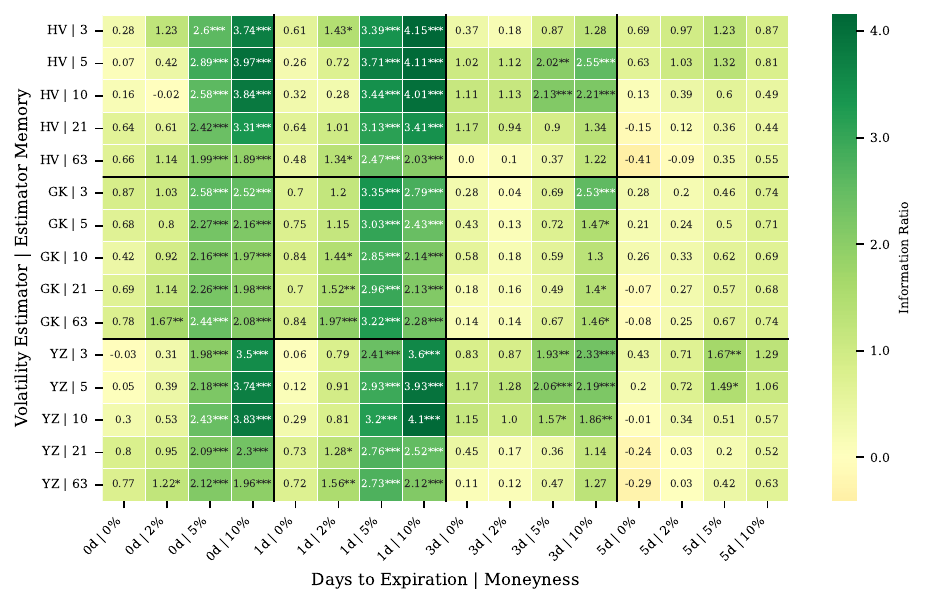}
\end{center}
\footnotesize{Note: Heatmap shows information ratios of short put strategies using three volatility estimators: historical (HV), Garman–Klass (GK), and Yang–Zhang (YZ). Estimator memory is the number of days in the volatility lookback window. *, **, and *** denote statistical significance at the 10\%, 5\%, and 1\% levels, respectively, from the PSR test against the benchmark B\&H strategy.}
\end{figure}

\begin{figure}[!ht]
\caption{\label{fig:kelly-arc} Annualized Returns of Short Put Strategies with Kelly Criterion Sizing}
\begin{center}
\includegraphics[width=0.75\columnwidth]{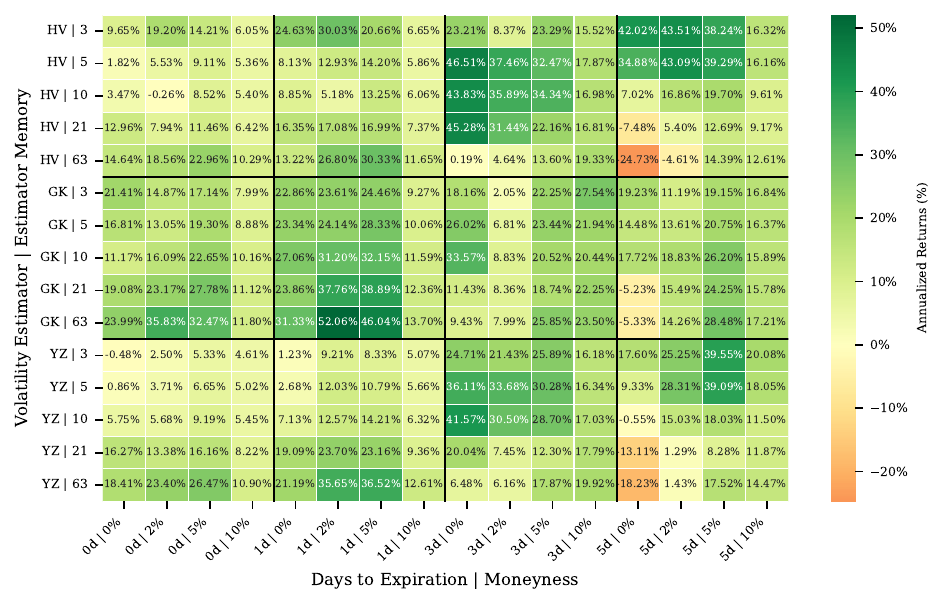}
\end{center}
\footnotesize{Note: Heatmap shows annualized returns of short put strategies using three volatility estimators: historical (HV), Garman–Klass (GK), and Yang–Zhang (YZ). Estimator memory is the number of days in the volatility lookback window. *, **, and *** denote statistical significance at the 10\%, 5\%, and 1\% levels, respectively, from the PSR test against the benchmark B\&H strategy.}
\end{figure}

Figures~\ref{fig:kelly-ir} and~\ref{fig:kelly-arc} display information ratios and annualized returns, respectively. Strategy performance is most sensitive to DTE and moneyness, while the choice of volatility estimator and lookback window has more subtle effects.

Moneyness is the most decisive driver of performance. Strategies using far OTM options (10\%) consistently yield positive and stable returns across DTEs, often exceeding 10–15\% annually. These strategies also achieve the highest information ratios—frequently above 3–4—and exhibit minimal drawdowns. In contrast, near-the-money options (0\%) display high return variability and severe drawdowns, particularly at longer DTEs (often exceeding 60–80\%).

DTE also has a substantial effect. Short-dated options (0 and 1 DTE) deliver the most favorable performance, particularly in combination with 5\%–10\% OTM. Medium-term options (3 DTE) perform well at higher moneyness levels, while longer-dated options (5 DTE) tend to underperform, especially near-the-money.

Volatility estimator choice has a more nuanced impact. The Garman-Klass estimator generally produces higher returns across most DTE/OTM configurations, but historical volatility and the Yang-Zhang estimator tend to outperform the Garman-Klass estimator when compared using the information ratio. The Yang-Zhang estimator performs better at longer DTEs (3d–5d), with more stable returns. Historical volatility shows inconsistent results, with both strong and weak performance depending on parameter combinations.

Estimator memory interacts with both estimator type and moneyness. Shorter windows (3–10 days) tend to amplify return volatility, occasionally boosting performance in far OTM strategies. Longer windows (21–63 days) provide greater stability and tend to favor the Garman-Klass estimator. Yang-Zhang and historical volatility benefit more from shorter windows in high OTM scenarios.

Overall, the most robust performance is observed for short-dated (0–1 DTE), far OTM (5–10\%) strategies. These configurations consistently yield the highest information ratios and the lowest drawdowns across all estimators. Among volatility estimators, the Garman-Klass variant with a 63-day window emerges as the top performer for 0 DTE options, while Yang-Zhang is more effective in 3d strategies. The highest returns are observed for strategies using 1DTE 2\% OTM options paired with the Garman-Klass estimator.

To sum up, the analysis reveals that moneyness and option maturity are the dominant factors driving strategy performance. Far OTM options (5-10\%) with short expiration windows (0-1 DTE) consistently demonstrate superior risk-adjusted returns and dramatically reduced drawdowns across all volatility estimators. The optimal parameter combination appears to be 0 or 1 DTE, 5\%-10\% OTM, and the Yang-Zhang estimator with 5-10 days of lookback window or the Garman-Klass with a 63-day lookback window. This region of the parameter space maximizes the trade‑off between annualized return, volatility, and drawdown, yielding high information ratios, annualized returns in the 20–25\% range, and controlled tail risk.

\subsection{Strategies with VIX-Rank Sizing}
\begin{figure}[!ht]
\caption{\label{fig:vix-ir} Information Ratios of Short Put Strategies with VIX-Rank Sizing}
\begin{center}
\includegraphics[width=0.75\columnwidth]{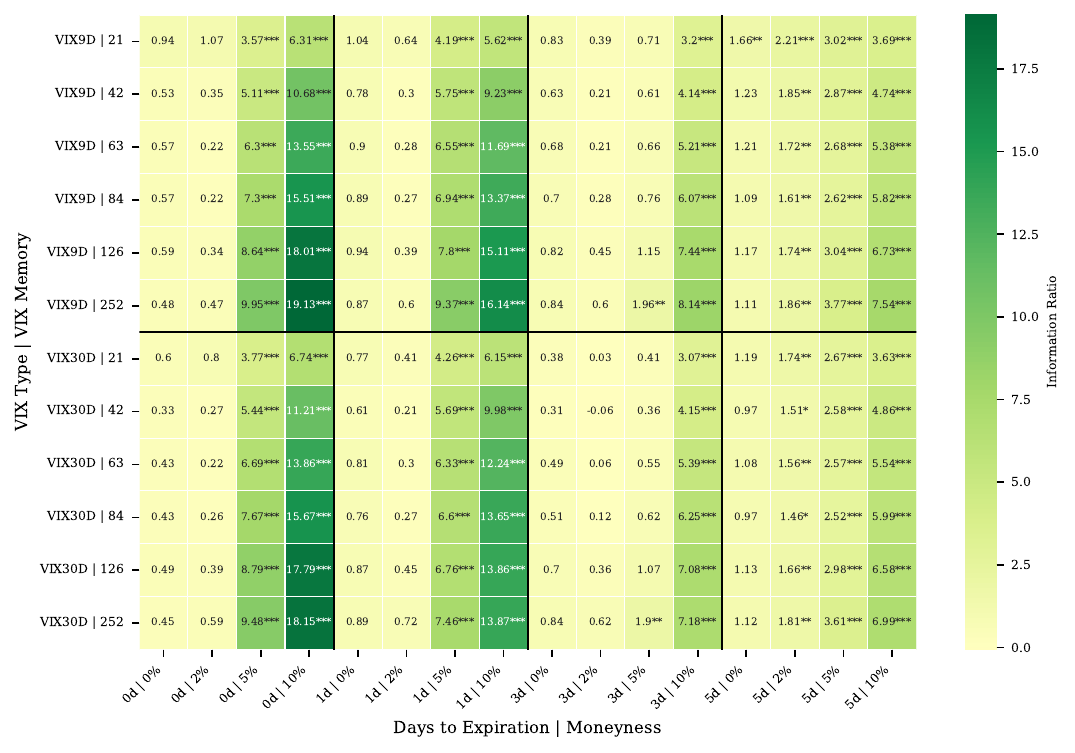}
\end{center}
\footnotesize{Note: Heatmap shows information ratios of short put strategies using two VIX measures: VIX9D and VIX30D. Memory indicates the number of days used to compute the VIX percentile rank. *, **, and *** denote statistical significance at the 10\%, 5\%, and 1\% levels, respectively, from the PSR test against the benchmark B\&H strategy.}
\end{figure}

\begin{figure}[!ht]
\caption{\label{fig:vix-arc} Annualized Returns of Short Put Strategies with VIX-Rank Sizing}
\begin{center}
\includegraphics[width=0.75\columnwidth]{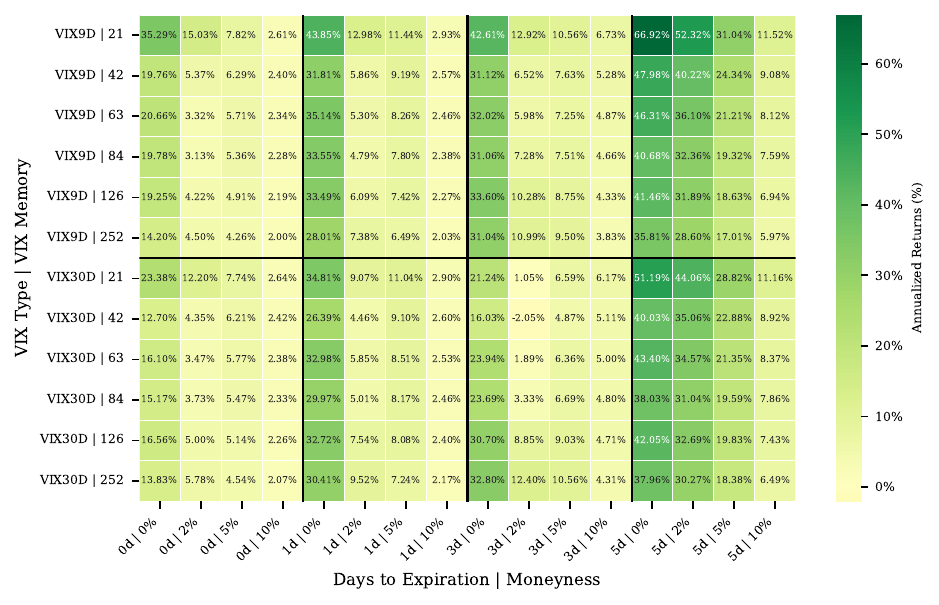}
\end{center}
\footnotesize{Note: Heatmap shows annualized returns of short put strategies using two VIX measures: VIX9D and VIX30D. Memory indicates the number of days used to compute the VIX percentile rank. *, **, and *** denote statistical significance at the 10\%, 5\%, and 1\% levels, respectively, from the PSR test against the benchmark B\&H strategy.}
\end{figure}

A clear and consistent pattern emerges in the relationship between moneyness and performance. As presented on Figure \ref{fig:vix-ir}, information ratios (IRs) increase with higher OTM levels across all DTE values and both VIX estimators. In contrast, as presented on Figure \ref{fig:vix-arc}, raw returns exhibit a non-linear decline as options move further out-of-the-money. The most pronounced drop in returns occurs between 0\% and 2\% OTM, suggesting a premium associated with precisely at-the-money options that is not linearly related to moneyness. While deep OTM options (5\%–10\%) yield lower absolute returns, they consistently deliver superior risk-adjusted performance, with IRs peaking at 10\% OTM.

The relationship between DTE and returns follows a clear upward trend: longer DTEs generate higher raw returns. This pattern is consistent across all memory windows and both VIX estimators. For example, using VIX9D with a 21-day memory at 0\% OTM, annualized returns nearly double from 35.29\% (0d) to 66.92\% (5d). However, this performance gain comes at the expense of stability—information ratios decline as DTE increases, indicating that higher returns are accompanied by increased volatility. Thus, shorter-dated strategies (particularly 0d DTE) consistently exhibit the highest IRs, especially for deep OTM options.

Regarding VIX type, VIX9D generally outperforms VIX30D in terms of both returns and IRs, particularly at shorter memory lengths and longer DTEs. As the memory window increases, the performance gap narrows, suggesting diminishing sensitivity to estimator choice with longer lookback periods.

Memory length effects vary by configuration. Medium-length windows (42–84 days) provide balanced performance across most settings. Shorter memory (21–42 days) favors ATM options with longer DTEs, while longer memory windows (126–252 days) support higher IRs in short-dated, far OTM configurations. This suggests that recent volatility has stronger predictive power for raw returns, whereas longer-term volatility history stabilizes risk-adjusted performance, particularly for conservative strategies.

Across nearly all configurations, the combination of 0 DTE, 10\% OTM, VIX9D, and a long memory window delivers the highest information ratios. In contrast, the highest raw returns are typically found in 5 DTE, 0\% OTM strategies using VIX9D with short memory. These findings highlight a trade-off between maximizing absolute performance and optimizing risk-adjusted outcomes. Investors seeking risk efficiency should favor the former, those seeking maximum return the latter. For investors seeking a balanced trade-off between return and risk, strategies based on 5~DTE options appear particularly attractive. These configurations tend to deliver higher returns than shorter-dated alternatives, while maintaining a manageable risk profile. A promising combination in this regard involves 5\% OTM options, paired with the 9-day VIX and a short memory in the rank methodology. This setup benefits from enhanced responsiveness to recent market conditions while avoiding the high risk exposure associated with at-the-money or very short-dated options. The resulting strategy offers a compelling compromise between performance and stability.

Overall, the robustness of these patterns across various configurations suggests the presence of structural features in volatility markets, rather than mere statistical artifacts. The consistent outperformance of ATM options in terms of raw returns versus deep OTM options in IRs reflects a moneyness-dependent volatility risk premium, likely driven by investors overpaying for tail risk protection. Moreover, the superior predictive power of shorter memory windows for returns aligns with known volatility clustering phenomena, while the systematic relationship between DTE and return magnitude likely reflects 0DTE-specific pricing in the short end of the volatility term structure. The performance differential between VIX9D and VIX30D further supports the hypothesis that short-term volatility estimates contain more actionable information for short put strategies.

\subsection{Strategies with Kelly-VIX Sizing}
\begin{figure}[!ht]
\caption{\label{fig:kelly-vix-30-ir} Information Ratios of Short Put Strategies with Kelly-VIX-Rank Sizing with VIX30D}
\begin{center}
\includegraphics[width=0.75\columnwidth]{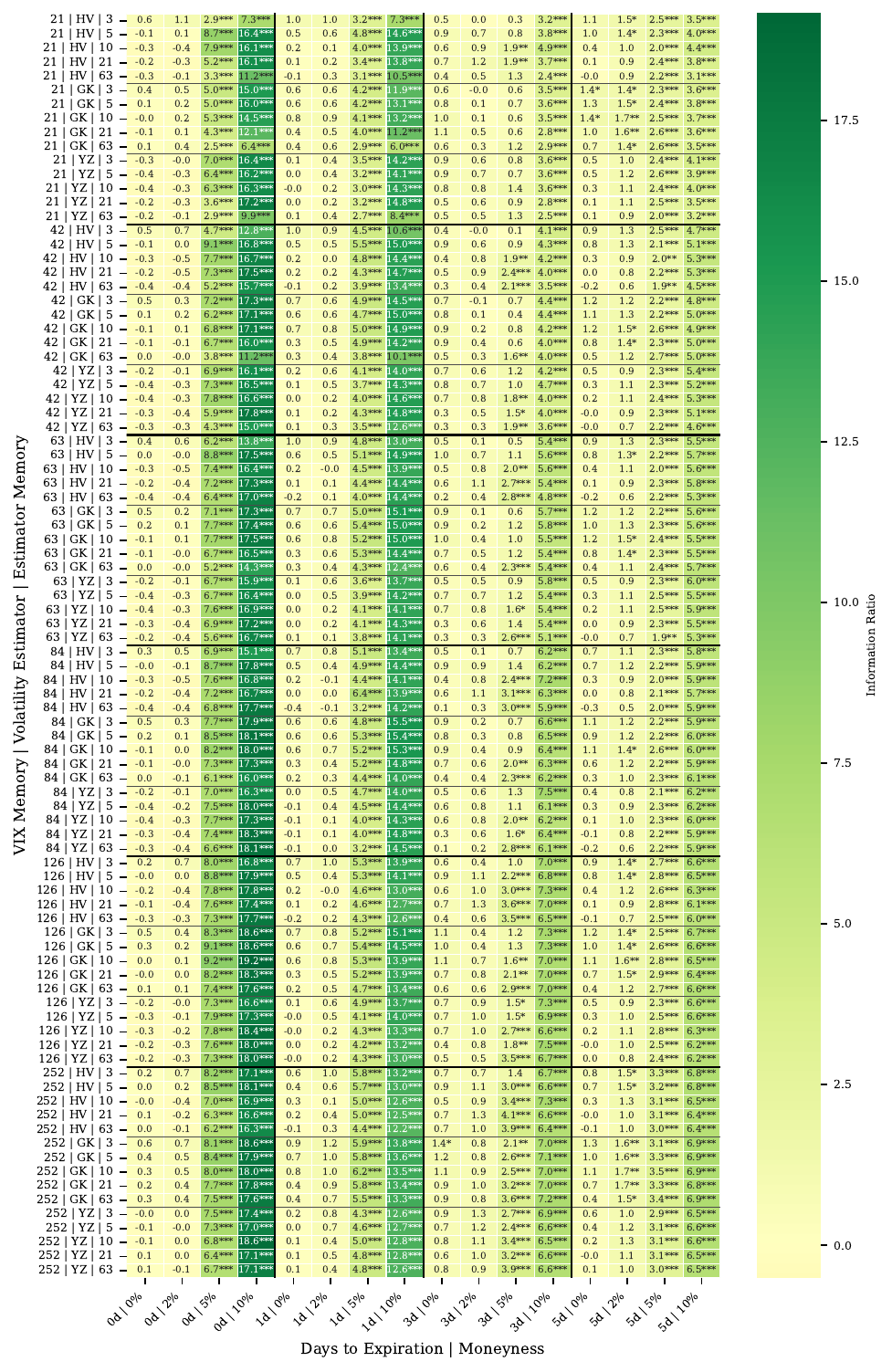}
\end{center}
\footnotesize{Note: Heatmap shows information ratios of short put strategies utilizing three volatility estimators: historical (HV), Garman–Klass (GK), and Yang–Zhang (YZ). Estimator memory is the number of days in the volatility lookback window. VIX memory indicates the number of days used to compute the VIX percentile rank. *, **, and *** denote statistical significance at the 10\%, 5\%, and 1\% levels, respectively, from the PSR test against the benchmark B\&H strategy.}
\end{figure}

\begin{figure}[!ht]
\caption{\label{fig:kelly-vix-30-arc} Annualized Returns of Short Put Strategies with Kelly-VIX-Rank Sizing with VIX30D}
\begin{center}
\includegraphics[width=0.75\columnwidth]{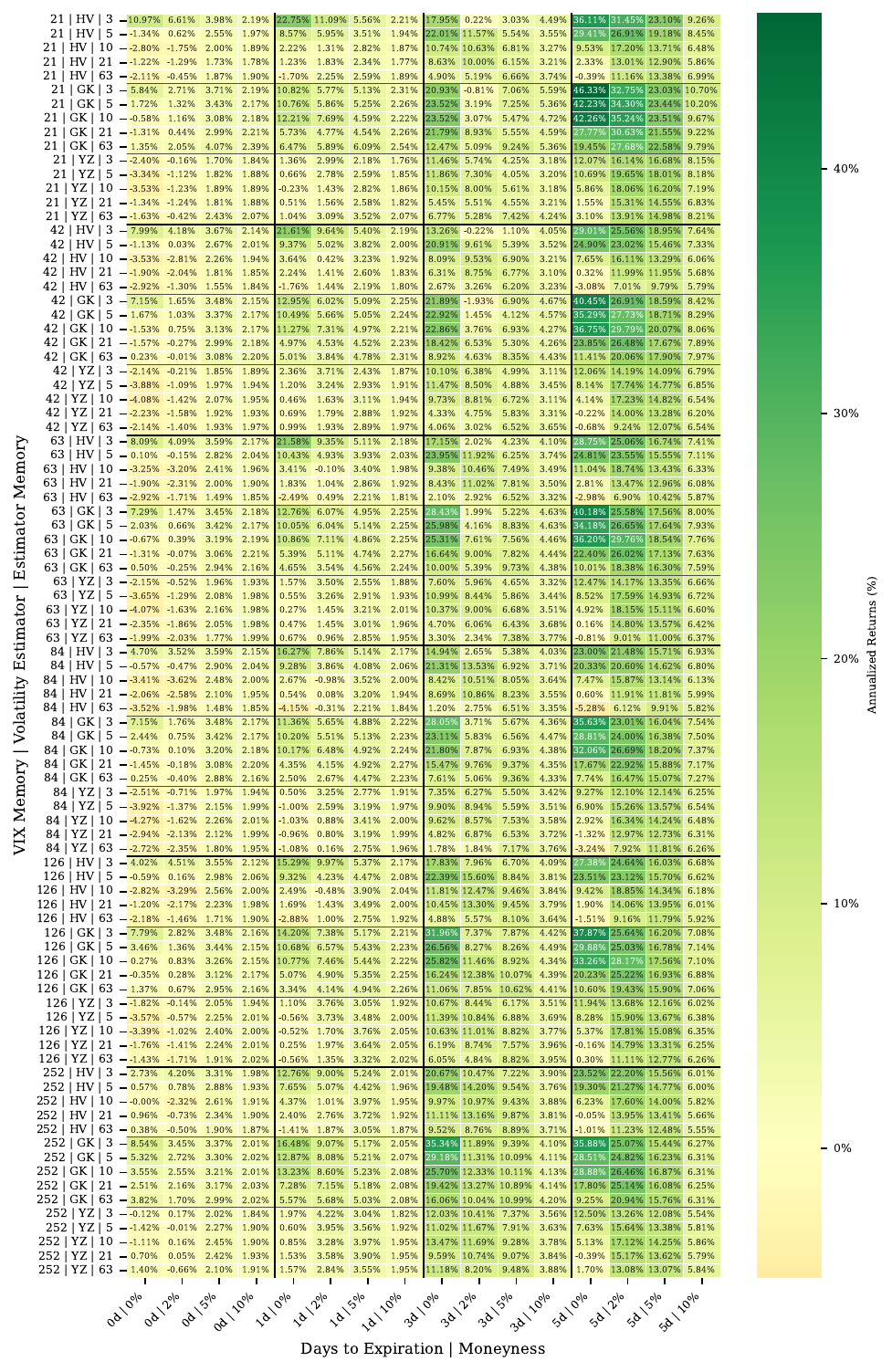}
\end{center}
\footnotesize{Note: Heatmap shows annualized returns of short put strategies utilizing three volatility estimators: historical (HV), Garman–Klass (GK), and Yang–Zhang (YZ). Estimator memory is the number of days in the volatility lookback window. VIX memory indicates the number of days used to compute the VIX percentile rank. *, **, and *** denote statistical significance at the 10\%, 5\%, and 1\% levels, respectively, from the PSR test against the benchmark B\&H strategy.}
\end{figure}

\begin{figure}[!ht]
\caption{\label{fig:kelly-vix-9-ir} Information Ratios of Short Put Strategies with Kelly-VIX-Rank Sizing with VIX9D}
\begin{center}
\includegraphics[width=0.75\columnwidth]{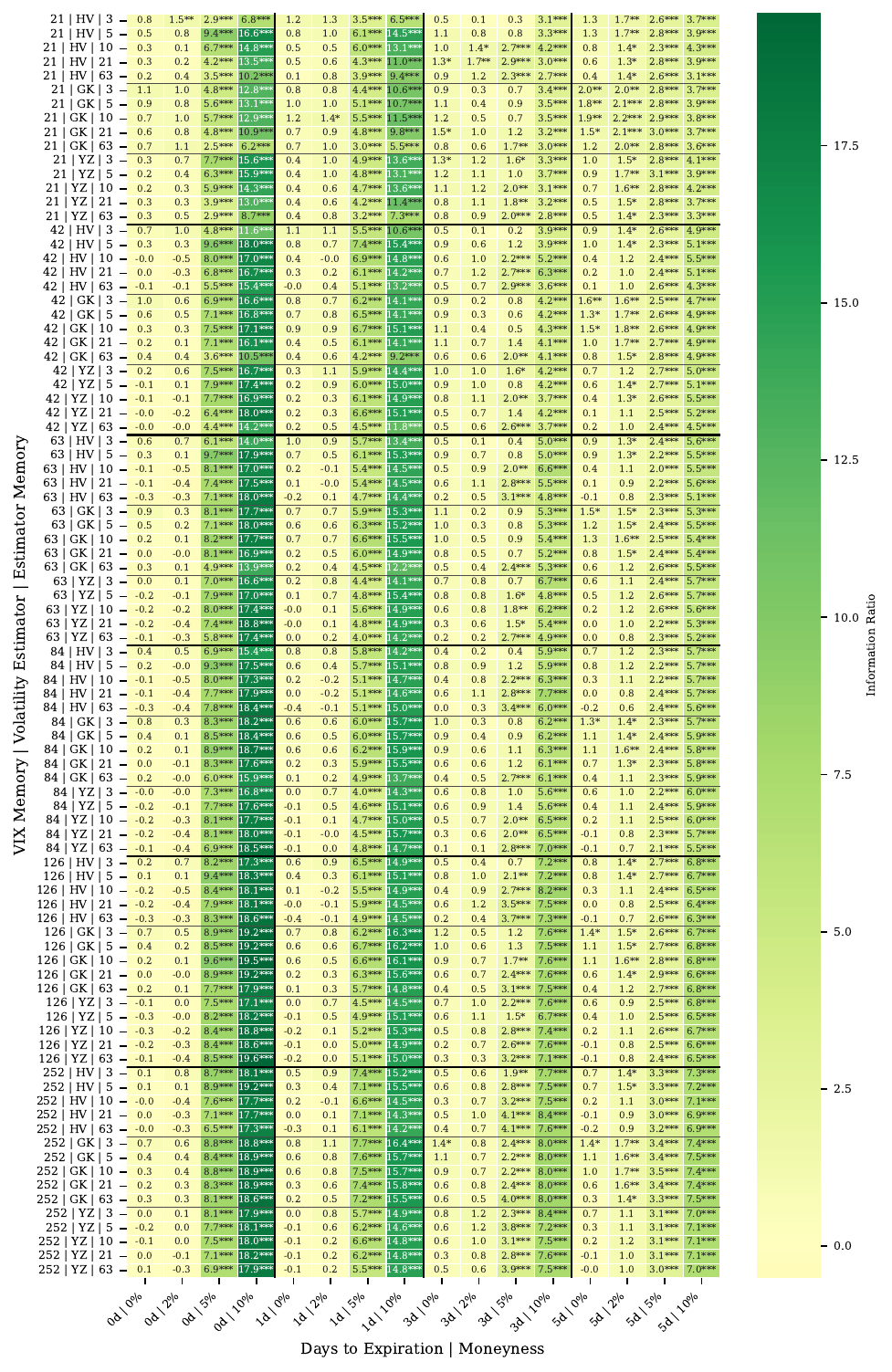}
\end{center}
\footnotesize{Note: Heatmap shows information ratios of short put strategies utilizing three volatility estimators: historical (HV), Garman–Klass (GK), and Yang–Zhang (YZ). Estimator memory is the number of days in the volatility lookback window. VIX memory indicates the number of days used to compute the VIX percentile rank. *, **, and *** denote statistical significance at the 10\%, 5\%, and 1\% levels, respectively, from the PSR test against the benchmark B\&H strategy.}
\end{figure}

\begin{figure}[!ht]
\caption{\label{fig:kelly-vix-9-arc} Annualized Returns of Short Put Strategies with Kelly-VIX-Rank Sizing with VIX9D}
\begin{center}
\includegraphics[width=0.75\columnwidth]{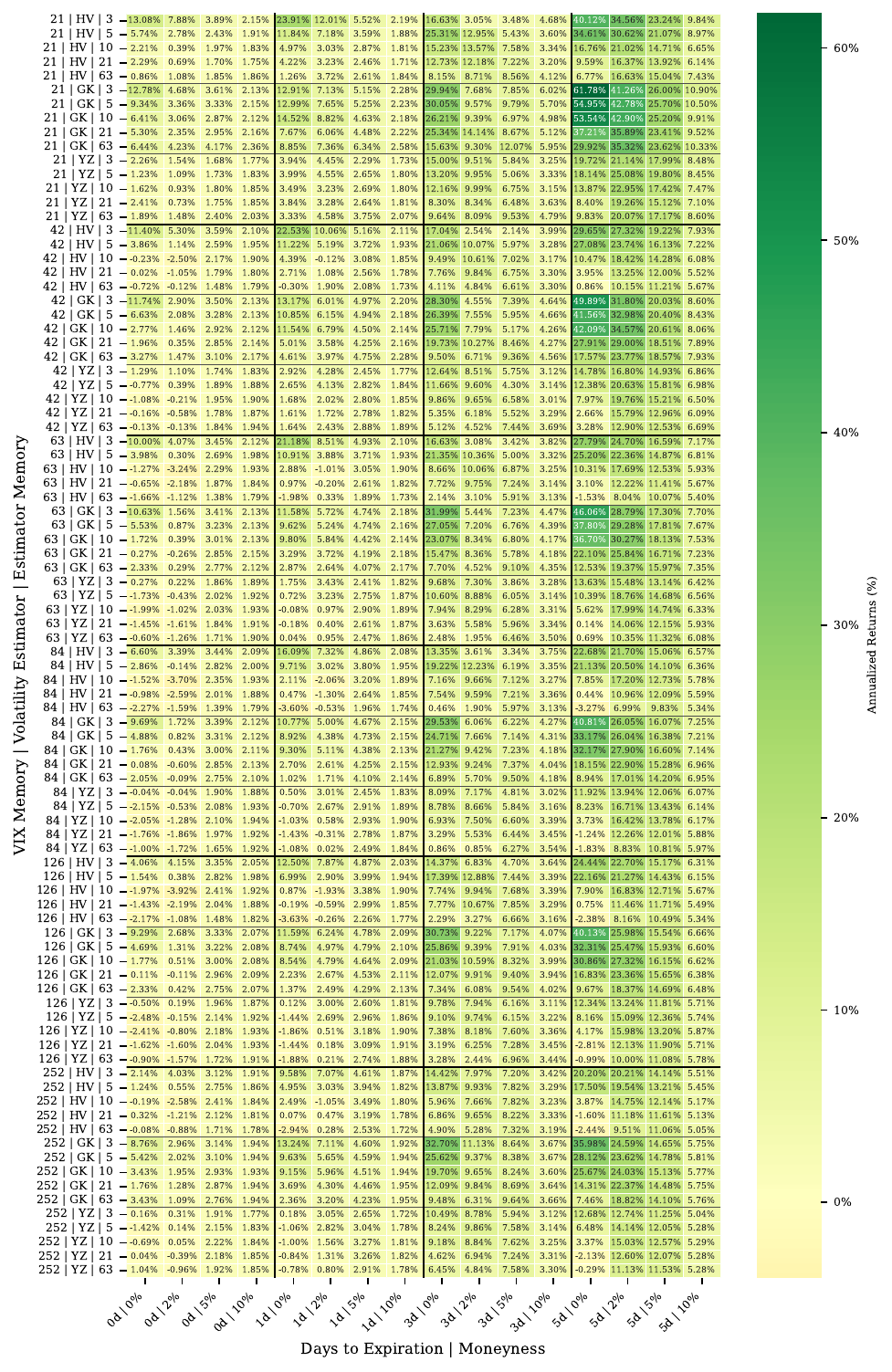}
\end{center}
\footnotesize{Note: Heatmap shows annualized returns of short put strategies utilizing three volatility estimators: historical (HV), Garman–Klass (GK), and Yang–Zhang (YZ). Estimator memory is the number of days in the volatility lookback window. VIX memory indicates the number of days used to compute the VIX percentile rank. *, **, and *** denote statistical significance at the 10\%, 5\%, and 1\% levels, respectively, from the PSR test against the benchmark B\&H strategy.}
\end{figure}

The heatmaps describing the performance of the Kelly-VIX combination sizing-based strategies are divided by VIX type to facilitate easier comparison and walk through results. Figures \ref{fig:kelly-vix-30-ir} and \ref{fig:kelly-vix-30-arc} provide an overview of annualized returns and information ratios of strategies with the VIX30D index, while Figures \ref{fig:kelly-vix-9-ir} and \ref{fig:kelly-vix-9-arc} summarize those metrics respectively for the VIX9D strategies.

At-the-money options consistently deliver the highest raw returns across most configurations. As moneyness increases to 2\%, 5\%, and 10\% OTM, returns decline in a generally monotonic fashion. The highest annualized returns, reaching approximately 40\%-45\%, are observed in strategies utilizing options with five days to expiration (5~DTE), at 0\%~OTM, combined with the Garman-Klass volatility estimator and short memory parameters (3 to 5 days). In contrast, the highest information ratios (IRs) are achieved with 0~DTE options, again under the Garman-Klass and Yang-Zhang estimators, and also using short memory.

While ATM configurations offer superior nominal returns, they are associated with elevated volatility, which translates into relatively low risk-adjusted performance across all maturities. Moving slightly out-of-the-money to 2\% leads to a substantial drop in returns with a mediocre improvement in IRs. However, strategies at 5\% and 10\% OTM, although providing lower absolute returns, yield markedly better risk-adjusted outcomes, indicating more effective control of downside risk. Most parameter configurations exhibit statistically significant alpha relative to the benchmark, underscoring the robustness of the observed effects.

The memory horizon of the VIX estimator also plays an important role. Longer memory settings, such as 126 to 252 trading days, tend to produce more stable volatility estimates and contribute to improved risk-adjusted returns. Conversely, shorter memory values (3 to 10 days) typically result in higher raw returns, reflecting increased responsiveness to short-term volatility and a greater exposure to market risk. This trade-off is consistent with the idea that recent data carries more predictive power for short-dated options, while longer-term volatility smoothing provides superior risk management.

In comparing volatility estimators, both the Garman-Klass and Yang-Zhang approaches consistently outperform the simpler historical volatility measure. Their ability to incorporate intraday price dynamics and account for discontinuities in opening prices appears to enhance the accuracy of volatility forecasts, thereby improving both raw and risk-adjusted performance. 

Across most configurations, short memory values in the volatility estimator (especially 3-day and 5-day windows) tend to dominate in terms of overall strategy performance. Medium memory values exhibit moderate effectiveness, while longer horizons generally underperform, likely due to the dilution of the relevant signal by older market data. These findings suggest that options-based strategies benefit substantially from focusing on recent market conditions when estimating volatility.

The standard deviation of returns increases as the strategy moves closer to ATM strikes, reflecting the higher risk and larger premium exposure inherent in such positions. The most severe maximum drawdowns are observed in 0\% OTM configurations, particularly those employing shorter DTEs, reinforcing the elevated risk associated with high-return strategies.

Overall, the consistency of performance across the parameter space indicates that these relationships are statistically robust and not the result of random variation. Optimal configurations vary depending on the target performance metric. For maximizing raw returns, the most favorable combination includes 5~DTE options, 0\%~OTM, the Garman-Klass estimator with a 3-day or 5-day memory, and a 63-day VIX memory. For maximizing risk-adjusted returns, the best results are obtained with 10\%~OTM options, the Garman-Klass estimator with a 5-day or 10-day memory, and a 126-day VIX memory. A more balanced trade-off between return and risk is achieved using 3~DTE, 5\%~OTM, the Garman-Klass estimator with 5-day memory, and a 252-day VIX memory, or similarly 5~DTE, 5\% OTM options paired with the Garman-Klass estimator with shorter memory of both the estimator and the VIX-rank. This configuration produces annualized returns in the range of 10\%--11\%, accompanied by statistically significant information ratios around 3.

In case of VIX9D, similar conclusions as to VIX30D can be drawn. Returns tend to increase with longer days‐to‐expiration (DTE), whereas information ratios (IRs) exhibit the opposite behavior: the strongest IRs occur at very short horizons (0 and 1 DTE), with still respectable, but noticeably lower, IRs for 3 and 5 DTE. Options with greater moneyness (further out‐of‐the‐money) deliver more stable performance and higher IRs, though at the cost of reduced raw returns.

Although the choice of volatility estimator influences results, no single estimator uniformly dominates across all parameter regimes. The Garman–Klass method typically outperforms simple historical volatility for short‐dated options, while the Yang–Zhang estimator excels in specific regions, especially when paired with short memory windows. Historical volatility generally underperforms both GK and YZ across most configurations.

Short memory spans (3–5 trading days) in volatility estimation often yield higher returns for short‐dated options, whereas longer spans (21–63 days) enhance stability but reduce absolute returns. The optimal memory horizon, however, depends on the combination of DTE and moneyness under consideration.

A similar trade‐off arises with VIX memory. Short VIX lookbacks improve returns for near‐term, near‐the‐money options but increase performance volatility, while long VIX horizons produce steadier yet lower returns. 

A comparison between the 9-day VIX and the standard 30-day VIX reveals that using VIX9D generally leads to superior performance in the context of this short-dated put-writing strategy. Specifically, VIX9D tends to deliver higher returns while maintaining comparable or even improved risk-adjusted performance metrics. Although the overall differences are moderate, the shorter horizon of VIX9D appears better aligned with the time frame of the options employed in the strategy, particularly for near-expiration positions. This suggests that the 30-day VIX may incorporate excessive noise or irrelevant longer-term expectations, which dilutes its utility for short-term volatility forecasting and risk management. Accordingly, VIX9D emerges as a more informative and responsive proxy for implied volatility in close-to-maturity option selling strategies.

Taken together, these results indicate that the put‐writing strategy effectively captures the volatility risk premium. The superior performance of very short‐dated options is consistent with known biases in short‐term volatility forecasting. Moreover, the outperformance of Garman-Klass and Yang-Zhang estimators underscores the value of using open‐high‐low‐close price information rather than relying solely on close‐to‐close returns, particularly for short‐horizon volatility prediction. Finally, the strategy benefits from the nonlinear time decay of the option premium, which accelerates as expiration approaches.

\section{Out-of-Sample Results}
The out-of-sample testing of the parameterized option-based investment strategies presented in Table \ref{tab:oos-results} reveals substantial variation in performance across different configurations, with several approaches achieving superior risk-adjusted returns compared to traditional benchmarks. Performance evaluation against the Buy-and-Hold (B\&H) strategy and the CBOE PutWrite Index (PUT) provides essential context. While absolute returns vary considerably, the strategies generally offer improved risk management characteristics, notably lower maximum drawdowns and enhanced information ratios.

The B\&H benchmark achieved an annualized return of 24\% with a volatility of 12.63\%, resulting in an information ratio of 1.9, an exceptionally strong outcome. The PUT index delivered 17.93\% annualized returns with significantly lower volatility of 6.52\%, producing an outstanding information ratio of 2.75. Notably, the year 2024 was marked by a strong bull market, which drove down put option premiums due to reduced perceived downside risk, thereby affecting the pricing and performance of volatility-selling strategies.

Strategies based on the Kelly criterion demonstrated consistent and robust risk control, particularly in 0 and 1~DTE configurations. These strategies maintained low maximum drawdowns across the test period, albeit with more modest absolute returns ranging from 2.53\% to 17.24\%, depending on the specific parameters. Among these, the 1~DTE and 5\% OTM configuration yielded the highest returns (14.35\%--17.24\%) with volatility around 8.5\%, significantly below the B\&H benchmark. The choice of volatility estimator also influenced performance, with the Garman–Klass (GK) estimator using a 63-day memory producing the highest returns, outperforming the Yang–Zhang (YZ) estimator under similar conditions.

The VIX-rank-based strategies exhibited the widest dispersion in outcomes, with annualized returns ranging from 0.95\% to an exceptional 52.77\%. This variation largely stemmed from the interaction between DTE and OTM parameters. The most aggressive configurations—5~DTE, at-the-money options, using VIX9D with a 21-day memory—achieved the highest return (52.77\%) with 21.59\% volatility and a maximum drawdown of 9.91\%. While volatility was significantly higher than benchmark levels, the resulting information ratio still exceeded that of the B\&H strategy, though it remained below the PUT index.

VIX memory length exerted a pronounced effect on both performance and stability. Longer memory periods (126--252 days), combined with conservative 10\% OTM positioning, yielded consistent but modest returns with exceptionally low volatility and strong information ratios. These configurations represent highly stable but low-return profiles. In contrast, shorter memory periods (21--42 days) supported more aggressive positioning, capable of capturing larger market moves at the cost of increased risk. A clear pattern emerges in the 5~DTE, 5\% OTM configurations: returns decrease from 6.07\% to 5.35\% as VIX memory increases from 21 to 63 days, indicating that longer memory horizons may reduce responsiveness to recent volatility shifts.

Hybrid Kelly–VIX strategies achieved annualized returns ranging from 0.90\% to 23.13\%, with the best results attained using VIX30D with a 63-day memory and short-memory GK volatility estimation. These hybrids effectively mitigated tail risk while capturing upside potential, with maximum drawdowns generally contained below 11\% even in aggressive configurations.

The interaction between VIX tenor (9-day vs.\ 30-day) and memory parameters offers additional insight into volatility regime detection. While VIX30D-based strategies achieved slightly higher peak performance (23.13\%) than VIX9D counterparts (22.11\%) in aggressive setups (5~DTE, ATM), this advantage diminishes in more conservative settings where the two measures yield comparable results. Estimator memory length plays a critical role, with short memory spans (3--5 days) consistently enhancing performance in fast-changing environments.

Cross-strategy analysis identifies several key parameter relationships that inform optimal configuration. Days-to-expiration emerges as a dominant factor, with 5~DTE strategies consistently outperforming alternatives. Moneyness analysis reaffirms earlier findings: further OTM options yield more stable but lower returns, which leads to underperformance in bull markets.

In summary, the risk-adjusted performance analysis suggests that, although challenging, it is possible to outperform traditional benchmarks even during strongly upward-trending markets. The comprehensive out-of-sample testing underscores that thoughtful and sophisticated parameterization of option-selling strategies can enhance risk-adjusted returns relative to benchmark strategies.

\begin{table}[!t]
\caption{Out-of-Sample Results of Short Put Strategies}
\label{tab:oos-results}
\resizebox{\columnwidth}{!}{%
\begin{tabular}{@{}ccccccccccccccc@{}}
\toprule
\textbf{SIZING} &
\textbf{DTE} &
  \textbf{\%OTM} &
  \textbf{VIX} &
  \makecell{\textbf{VIX}\\\textbf{MEM.}} &
  \textbf{EST.} &
  \makecell{\textbf{EST.}\\\textbf{MEM.}} &
  \makecell{\textbf{aRC}\\\textbf{(\%)}} &
  \makecell{\textbf{aSD}\\\textbf{(\%)}} &
  \makecell{\textbf{MD}\\\textbf{(\%)}} &
  \textbf{MLD} &
  \textbf{IR} &
  \makecell{\textbf{VaR}\\\textbf{(\%)}} &
  \makecell{\textbf{CVaR}\\\textbf{(\%)}} &
  \textbf{N} \\ \hline
  B\&H & - & - & - & - & - & - & 24.01 & 12.63 & 8.49 & 0.18 & 1.9 & 1.34 & 1.89 & 1 \\
  PUT & - & - & - & - & - & - & 17.93 & 6.52 & 4.96 & 0.07 & 2.75 & 0.5 &  0.99 & 1 \\
  \arrayrulecolor{black!50}\midrule
Kelly           & 0            & 5            & -            & -                   & YZ                 & 5                         & 7,54         & 5,91         & 0,00        & 0,00         & 1.28        & 0,00         & 0,00          & 236        \\
Kelly           & 0            & 5            & -            & -                   & YZ                 & 10                        & 7,73         & 5,91         & 0,00        & 0,00         & 1.31        & 0,00         & 0,00          & 240        \\
Kelly           & 0            & 10           & -            & -                   & YZ                 & 5                         & 2,55         & 0,57         & 0,00        & 0,00         & 4.44        & 0,01         & 0,00          & 245        \\
Kelly           & 0            & 10           & -            & -                   & YZ                 & 10                        & 2,53         & 0,57         & 0,00        & 0,00         & 4.41        & 0,01         & 0,00          & 250        \\
\arrayrulecolor{black!10}\midrule
Kelly           & 1            & 5            & -            & -                   & YZ                 & 5                         & 14,35        & 8,44         & 0,07        & 0,00         & 1.7         & 0,00         & 0,00          & 183        \\
Kelly           & 1            & 5            & -            & -                   & YZ                 & 10                        & 14,72        & 8,42         & 0,07        & 0,01         & 1.75        & 0,00         & -0,01         & 185        \\
Kelly           & 1            & 5            & -            & -                   & GK                 & 63                        & 17,24        & 8,47         & 0,07        & 0,00         & 2.03        & 0,00         & -0,01         & 195        \\
Kelly           & 1            & 10           & -            & -                   & YZ                 & 5                         & 7,35         & 4,57         & 0,00        & 0,00         & 1.61        & 0,00         & 0,00          & 192        \\
Kelly           & 1            & 10           & -            & -                   & YZ                 & 10                        & 7,42         & 4,57         & 0,00        & 0,00         & 1.62        & 0,00         & 0,00          & 193        \\
Kelly           & 1            & 10           & -            & -                   & GK                 & 63                        & 7,52         & 4,57         & 0,00        & 0,00         & 1.64        & 0,00         & 0,00          & 195        \\
\arrayrulecolor{black!50}\midrule
VIX-Rank        & 0            & 10           & VIX9D        & 126                 & -                  & -                         & 0,95         & 0,04         & 0,00         & 0,00          & 25.4     & 0,00         & 0,00          & 246        \\
VIX-Rank        & 0            & 10           & VIX9D        & 252                 & -                  & -                         & 1            & 0,04         & 0,00         & 0,00          & 25.23    & 0,00         & 0,00          & 245        \\
\arrayrulecolor{black!10}\midrule
VIX-Rank        & 5            & 0            & VIX9D        & 21                  & -                  & -                         & 52,77        & 21,59        & 9,91        & 0,09         & 2.44        & -2,11        & -3,96         & 178        \\
VIX-Rank        & 5            & 0            & VIX9D        & 42                  & -                  & -                         & 35,45        & 22,55        & 14,96       & 0,1          & 1.57        & -1,84        & -4,32         & 182        \\
VIX-Rank        & 5            & 5            & VIX9D        & 21                  & -                  & -                         & 6,07         & 1,29         & 0,91        & 0,08         & 4.7         & -0,02        & -0,18         & 178        \\
VIX-Rank        & 5            & 5            & VIX9D        & 42                  & -                  & -                         & 5,78         & 0,98         & 0,52        & 0,09         & 5.92        & -0,02        & -0,15         & 182        \\
VIX-Rank        & 5            & 5            & VIX9D        & 63                  & -                  & -                         & 5,35         & 0,94         & 0,59        & 0,13         & 5.69        & -0,02        & -0,14         & 187        \\
\arrayrulecolor{black!50}\midrule
Kelly-VIX       & 0            & 10           & VIX9D        & 126                 & GK                 & 10                        & 0,95         & 0,04         & 0,00         & 0,00          & 25.36    & 0,00         & 0,00          & 246        \\
Kelly-VIX       & 0            & 10           & VIX30D       & 126                 & GK                 & 10                        & 0,9          & 0,04         & 0,00         & 0,00          & 23.69    & 0,00         & 0,00          & 245        \\
\arrayrulecolor{black!10}\midrule
Kelly-VIX       & 3            & 5            & VIX9D        & 252                 & GK                 & 5                         & 2,38         & 2,11         & 1,58        & 0,25         & 1.13        & 0,00         & -0,23         & 148        \\
Kelly-VIX       & 3            & 5            & VIX30D       & 252                 & GK                 & 5                         & 2,4          & 2,11         & 1,62        & 0,27         & 1.14        & -0,01        & -0,24         & 148        \\
\arrayrulecolor{black!10}\midrule
Kelly-VIX       & 5            & 0            & VIX9D        & 63                  & GK                 & 3                         & 22,11        & 17,98        & 10,74       & 0,19         & 1.23        & -1,4         & -3,2          & 140        \\
Kelly-VIX       & 5            & 0            & VIX9D        & 63                  & GK                 & 5                         & 16,87        & 17,95        & 10,31       & 0,19         & 0.94        & -1,17        & -3,15         & 142        \\
Kelly-VIX       & 5            & 0            & VIX30D       & 63                  & GK                 & 3                         & 23,13        & 18,46        & 9,46        & 0,11         & 1.25        & -1,2         & -3,11         & 137        \\
Kelly-VIX       & 5            & 0            & VIX30D       & 63                  & GK                 & 5                         & 17,34        & 18,28        & 9,42        & 0,12         & 0.95        & -1,2         & -3,1          & 138        \\
Kelly-VIX       & 5            & 5            & VIX30D       & 21                  & GK                 & 5                         & 5,02         & 1,48         & 1,07        & 0,15         & 3.39        & -0,02        & -0,2          & 176        \\
Kelly-VIX       & 5            & 5            & VIX9D        & 21                  & GK                 & 5                         & 5,72         & 1,28         & 0,91        & 0,11         & 4.47        & -0,01        & -0,18         & 178 \\
\arrayrulecolor{black}\hline
\end{tabular}%
}
\source{This table presents results of strategies on the out-of-sample period from 01-01-2024 to 31-12-2024. B\&H sizing presents results of a benchmark buy-and-hold strategy. PUT sizing presents results of a benchmark strategy investing in the CBOE PUT index. Column N contains the total number of positions opened by the specified strategy. Moneyness (\%OTM) is expressed as a percentage of out-of-the-money. *, **, and *** denote statistical significance at the 10\%, 5\%, and 1\% levels, respectively, from the probabilistic Sharpe Ratio test against the benchmark buy-and-hold (B\&H) strategy. The VIX and estimator memory values are expressed in the number of days used to calculate the percentile rank of the most recent VIX value.}
\end{table}

\section{Conclusions}
This comprehensive study evaluates the performance characteristics of systematic put-writing strategies across multiple dimensions, focusing on three position sizing methodologies: the Kelly criterion, VIX-rank-based sizing, and a hybrid Kelly–VIX approach. The analysis spans variations in days-to-expiry (DTE), moneyness, volatility estimation methods, and memory window parameters, incorporating both in-sample optimization and rigorous out-of-sample validation.

The results underscore that moneyness and DTE are the primary determinants of strategy performance, consistently exerting greater influence than volatility estimator choice or memory length. Strategies employing far out-of-the-money (OTM) options, particularly 5\% to 10\% OTM, demonstrate superior risk-adjusted returns relative to at-the-money configurations, indicating the presence of a persistent volatility risk premium that can be effectively captured through informed parameter selection. In general, findings are consistent with previous literature, indicating that strategies closer to the ATM levels can provide higher returns, but come with elevated risk \citep{schwalbach_analysis_2018}.

The temporal structure of option exposure is equally decisive. Extremely short-dated options, especially 0–1~DTE, exhibit the highest information ratios and favorable drawdown profiles. This aligns with theoretical expectations around rapid time decay near expiration and the predictive power of short-horizon volatility estimates. However, the highest absolute returns were typically realized with 5~DTE options, suggesting that slightly longer-dated positions offer a better balance between risk and reward, while still enabling precise risk control.

Among volatility estimators, the Garman–Klass and Yang–Zhang methods outperform standard historical volatility. Their ability to incorporate intraday price action and overnight jumps significantly enhances short-term forecasting. The Garman–Klass estimator is most effective with longer memory windows (approximately 63 days), while Yang–Zhang performs best with shorter memories (5–10 days), offering flexibility depending on the strategy horizon. These results are also generally in line with the previous literature in the field, indicating good performance of the Garman-Klass estimator \citep{molnar_properties_2012}.

Comparative analysis between VIX9D and VIX30D reveals a consistent performance edge for the shorter-term measure. VIX9D-based strategies yield higher returns with equal or superior risk-adjusted metrics, suggesting that the 30-day VIX may incorporate excessive noise or lag for short-dated strategies. This insight has important implications for volatility forecasting and dynamic exposure calibration.

The out-of-sample testing period, marked by strong equity market gains throughout 2024, posed a challenging environment for volatility-selling strategies due to compressed put premiums. Nevertheless, several configurations achieved annualized returns of 14\%–23\% with materially lower volatility and drawdowns than the buy-and-hold benchmark, confirming the resilience of well-designed approaches under adverse conditions.

Hybrid Kelly–VIX strategies emerged as robust, balancing return generation with risk management. These strategies adaptively adjust position sizing based on theoretical allocation rules and prevailing market volatility conditions. Their consistent performance demonstrates the value of integrating multiple informational dimensions into the sizing process, particularly under regimes of suppressed volatility and sustained equity strength.

From a portfolio construction standpoint, systematic option strategies exhibit strong potential as complements or alternatives to traditional equity allocations. Their distinct risk profiles, lower correlation to conventional asset classes, and capacity to generate positive returns in various market environments position them as valuable diversification tools for institutional and sophisticated individual investors.

While the findings are compelling, several limitations merit attention. The study focuses exclusively on S\&P 500 index options; extending the analysis to alternative underlyings and international markets is a natural next step. Additionally, while the test period includes one year of data, it does not encompass all volatility regimes, particularly extreme dislocations. Future research should also assess sensitivity to varying cost assumptions, including varying slippage and market impact, and explore regime-switching models that dynamically adapt strategy parameters. Such enhancements could further refine risk-return profiles and improve robustness across market environments.

In summary, this research provides robust empirical evidence that carefully parameterized systematic put-writing strategies can generate attractive risk-adjusted returns across a range of market conditions. The identification of optimal configurations offers actionable insights for strategy implementation, while the results contribute meaningfully to the broader understanding of volatility risk premium dynamics. The consistent outperformance of sophisticated designs over naïve implementations highlights the importance of disciplined model development and reinforces the practical value of these strategies in modern investment portfolios.

\clearpage
\bibliography{hedged}

\clearpage
\appendix

\section{Short Put Strategies with Kelly Criterion Sizing}
\begin{table}[!h]
\caption{Results of Short Put Strategies with Kelly Criterion Sizing with 0 or 1 Days to Expiration}
\label{tab:kelly-results-0-1}
\resizebox{\columnwidth}{!}{%
\begin{tabular}{@{}cccccccccccc|cccccccccccc@{}}
\toprule
\textbf{DTE} &
  \textbf{\%OTM} &
  \textbf{EST.} &
  \makecell{\textbf{EST.}\\\textbf{MEM.}} &
  \makecell{\textbf{aRC}\\\textbf{(\%)}} &
  \makecell{\textbf{aSD}\\\textbf{(\%)}} &
  \makecell{\textbf{MD}\\\textbf{(\%)}} &
  \textbf{MLD} &
  \makecell{\textbf{IR}\\\textbf{(\%)}} &
  \makecell{\textbf{VaR}\\\textbf{(\%)}} &
  \makecell{\textbf{CVaR}\\\textbf{(\%)}} &
  \textbf{N} &
  \textbf{DTE} &
  \textbf{\%OTM} &
  \textbf{EST.} &
  \makecell{\textbf{EST.}\\\textbf{MEM.}} &
  \makecell{\textbf{aRC}\\\textbf{(\%)}} &
  \makecell{\textbf{aSD}\\\textbf{(\%)}} &
  \makecell{\textbf{MD}\\\textbf{(\%)}} &
  \textbf{MLD} &
  \textbf{IR} &
  \makecell{\textbf{VaR}\\\textbf{(\%)}} &
  \makecell{\textbf{CVaR}\\\textbf{(\%)}} &
  \textbf{N} \\ \midrule
0 & 0  & GK & 3  & 21,41 & 24,62 & 24,59 & 0,65 & 0.87    & -0,9  & -3,83 & 312  & 1 & 0  & GK & 3  & 22,86 & 32,43 & 53,31 & 1,24 & 0.7     & -2,25 & -5,71 & 399 \\
0 & 0  & GK & 5  & 16,81 & 24,6  & 24,14 & 1,15 & 0.68    & -0,81 & -3,83 & 276  & 1 & 0  & GK & 5  & 23,34 & 31,18 & 42,51 & 0,74 & 0.75    & -1,68 & -5,42 & 369 \\
0 & 0  & GK & 10 & 11,17 & 26,57 & 30,29 & 1,17 & 0.42    & -0,91 & -4,17 & 256  & 1 & 0  & GK & 10 & 27,06 & 32,18 & 33,68 & 0,53 & 0.84    & -1,49 & -5,39 & 326 \\
0 & 0  & GK & 21 & 19,08 & 27,71 & 26,51 & 0,66 & 0.69    & -1,18 & -4,28 & 244  & 1 & 0  & GK & 21 & 23,86 & 34,09 & 35,17 & 0,74 & 0.7     & -2,48 & -5,78 & 321 \\
0 & 0  & GK & 63 & 23,99 & 30,85 & 34,02 & 0,96 & 0.78    & -0,99 & -4,53 & 234  & 1 & 0  & GK & 63 & 31,33 & 37,2  & 38,39 & 1,12 & 0.84    & -2,23 & -6,02 & 301 \\
\arrayrulecolor{black!10}\midrule
0 & 0  & HV & 3  & 9,65  & 34,63 & 49,54 & 2,7  & 0.28    & -3,43 & -6,42 & 443  & 1 & 0  & HV & 3  & 24,63 & 40,3  & 50,32 & 1,36 & 0.61    & -3,9  & -7,46 & 487 \\
0 & 0  & HV & 5  & 1,82  & 25,48 & 39,57 & 2,97 & 0.07    & -1,21 & -4,56 & 282  & 1 & 0  & HV & 5  & 8,13  & 31,43 & 35,79 & 2,71 & 0.26    & -2,44 & -5,83 & 342 \\
0 & 0  & HV & 10 & 3,47  & 21,9  & 30,51 & 2,1  & 0.16    & -0,67 & -3,31 & 189  & 1 & 0  & HV & 10 & 8,85  & 27,33 & 28,93 & 2,02 & 0.32    & -1,03 & -4,19 & 239 \\
0 & 0  & HV & 21 & 12,96 & 20,1  & 26,4  & 1,07 & 0.64    & -0,00   & -0,13 & 139  & 1 & 0  & HV & 21 & 16,35 & 25,36 & 29,97 & 1,18 & 0.64    & -0,5  & -3,47 & 183 \\
0 & 0  & HV & 63 & 14,64 & 22,19 & 25,85 & 1,25 & 0.66    & -0,00   & -0,13 & 128  & 1 & 0  & HV & 63 & 13,22 & 27,35 & 33,15 & 1,28 & 0.48    & -0,45 & -3,6  & 162 \\
\arrayrulecolor{black!10}\midrule
0 & 0  & YZ & 3  & -0,48 & 17,07 & 27,99 & 1,36 & -0.03   & -0,00  & -0,11 & 98   & 1 & 0  & YZ & 3  & 1,23  & 21,32 & 32,27 & 1,14 & 0.06    & -0,1  & -2,85 & 138 \\
0 & 0  & YZ & 5  & 0,86  & 17,67 & 27,02 & 1,36 & 0.05    & -0,00  & -0,11 & 107  & 1 & 0  & YZ & 5  & 2,68  & 23,21 & 32,67 & 0,56 & 0.12    & -0,08 & -3,3  & 156 \\
0 & 0  & YZ & 10 & 5,75  & 19,25 & 28,91 & 1,29 & 0.3     & -0,00  & -0,13 & 123  & 1 & 0  & YZ & 10 & 7,13  & 24,8  & 31,74 & 0,53 & 0.29    & -0,22 & -3,6  & 173 \\
0 & 0  & YZ & 21 & 16,27 & 20,3  & 24,93 & 1,09 & 0.8     & -0,01 & -2,37 & 149  & 1 & 0  & YZ & 21 & 19,09 & 26,09 & 28,92 & 0,51 & 0.73    & -0,57 & -3,64 & 189 \\
0 & 0  & YZ & 63 & 18,41 & 23,91 & 24,19 & 1,07 & 0.77    & -0,02 & -2,89 & 156  & 1 & 0  & YZ & 63 & 21,19 & 29,48 & 34,02 & 0,93 & 0.72    & -0,65 & -4,22 & 207 \\
\arrayrulecolor{black!50}\midrule
0 & 2  & GK & 3  & 14,87 & 14,45 & 17,13 & 1,35 & 1.03    & -0,00  & -0,18 & 689  & 1 & 2  & GK & 3  & 23,61 & 19,63 & 23,36 & 0,92 & 1.2     & -0,4  & -2,9  & 752 \\
0 & 2  & GK & 5  & 13,05 & 16,28 & 16,07 & 0,87 & 0.8     & -0,00  & -0,19 & 678  & 1 & 2  & GK & 5  & 24,14 & 20,9  & 18,3  & 0,43 & 1.15    & -0,48 & -2,91 & 723 \\
0 & 2  & GK & 10 & 16,09 & 17,48 & 17,8  & 0,7  & 0.92    & -0,00  & -0,18 & 638  & 1 & 2  & GK & 10 & 31,2  & 21,68 & 18,59 & 0,56 & 1.44*   & -0,26 & -2,7  & 697 \\
0 & 2  & GK & 21 & 23,17 & 20,31 & 17,86 & 0,49 & 1.14    & -0,00  & -0,18 & 597  & 1 & 2  & GK & 21 & 37,76 & 24,88 & 20,51 & 0,43 & 1.52*   & -0,24 & -3,05 & 654 \\
0 & 2  & GK & 63 & 35,83 & 21,4  & 17,83 & 0,52 & 1.67**  & -0,00  & -0,14 & 493  & 1 & 2  & GK & 63 & 52,06 & 26,41 & 24,48 & 0,4  & 1.97*** & -0,34 & -3,07 & 577 \\
\arrayrulecolor{black!10}\midrule
0 & 2  & HV & 3  & 19,2  & 15,59 & 15,42 & 0,88 & 1.23    & -0,00  & -0,2  & 684  & 1 & 2  & HV & 3  & 30,03 & 20,94 & 20,54 & 0,74 & 1.43*   & -0,22 & -2,96 & 689 \\
0 & 2  & HV & 5  & 5,53  & 13,22 & 14,75 & 1,56 & 0.42    & -0,00  & -0,13 & 534  & 1 & 2  & HV & 5  & 12,93 & 17,96 & 22,39 & 1,42 & 0.72    & -0,00  & -2,42 & 585 \\
0 & 2  & HV & 10 & -0,26 & 13,99 & 23,51 & 4,27 & -0.02   & -0,00  & -0,11 & 407  & 1 & 2  & HV & 10 & 5,18  & 18,35 & 23,58 & 2,74 & 0.28    & -0,00  & -0,21 & 496 \\
0 & 2  & HV & 21 & 7,94  & 13,06 & 13,85 & 1,29 & 0.61    & -0,00  & -0,07 & 330  & 1 & 2  & HV & 21 & 17,08 & 16,92 & 17,36 & 1,2  & 1.01    & -0,00  & -0,13 & 425 \\
0 & 2  & HV & 63 & 18,56 & 16,28 & 14,55 & 1,29 & 1.14    & -0,00  & -0,06 & 219  & 1 & 2  & HV & 63 & 26,8  & 19,94 & 19,68 & 0,39 & 1.34*   & -0,00  & -0,11 & 308 \\
\arrayrulecolor{black!10}\midrule
0 & 2  & YZ & 3  & 2,5   & 8,12  & 13,93 & 1,29 & 0.31    & -0,00  & -0,02 & 248  & 1 & 2  & YZ & 3  & 9,21  & 11,66 & 14,65 & 0,33 & 0.79    & -0,00  & -0,06 & 335 \\
0 & 2  & YZ & 5  & 3,71  & 9,5   & 13,86 & 1,29 & 0.39    & -0,00  & -0,04 & 298  & 1 & 2  & YZ & 5  & 12,03 & 13,19 & 18,45 & 0,7  & 0.91    & -0,00  & -0,09 & 395 \\
0 & 2  & YZ & 10 & 5,68  & 10,69 & 13,8  & 1,29 & 0.53    & -0,00  & -0,05 & 337  & 1 & 2  & YZ & 10 & 12,57 & 15,46 & 18,47 & 0,7  & 0.81    & -0,00  & -0,13 & 439 \\
0 & 2  & YZ & 21 & 13,38 & 14,06 & 13,86 & 1,29 & 0.95    & -0,00  & -0,06 & 324  & 1 & 2  & YZ & 21 & 23,7  & 18,51 & 19,07 & 0,7  & 1.28*   & -0,00  & -0,14 & 424 \\
0 & 2  & YZ & 63 & 23,4  & 19,18 & 16,83 & 1,29 & 1.22*   & -0,00  & -0,08 & 269  & 1 & 2  & YZ & 63 & 35,65 & 22,82 & 20,55 & 0,39 & 1.56**  & -0,00  & -0,16 & 373 \\
\arrayrulecolor{black!50}\midrule
0 & 5  & GK & 3  & 17,14 & 6,64  & 0,52  & 0,13 & 2.58*** & -0,00  & -0,01 & 1005 & 1 & 5  & GK & 3  & 24,46 & 7,31  & 1,68  & 0,08 & 3.35*** & -0,00  & -0,06 & 922 \\
0 & 5  & GK & 5  & 19,3  & 8,49  & 0,28  & 0,13 & 2.27*** & -0,00  & -0,00  & 1018 & 1 & 5  & GK & 5  & 28,33 & 9,35  & 1,68  & 0,04 & 3.03*** & -0,00  & -0,04 & 930 \\
0 & 5  & GK & 10 & 22,65 & 10,49 & 0,21  & 0,02 & 2.16*** & -0,00  & -0,00  & 1022 & 1 & 5  & GK & 10 & 32,15 & 11,28 & 1,68  & 0,03 & 2.85*** & -0,00  & -0,04 & 932 \\
0 & 5  & GK & 21 & 27,78 & 12,3  & 0,25  & 0,02 & 2.26*** & -0,00  & -0,00  & 1019 & 1 & 5  & GK & 21 & 38,89 & 13,14 & 1,68  & 0,05 & 2.96*** & -0,00  & -0,04 & 931 \\
0 & 5  & GK & 63 & 32,47 & 13,29 & 0,07  & 0,01 & 2.44*** & -0,00  & -0,00  & 1037 & 1 & 5  & GK & 63 & 46,04 & 14,31 & 1,68  & 0,06 & 3.22*** & -0,00  & -0,03 & 911 \\
\arrayrulecolor{black!10}\midrule
0 & 5  & HV & 3  & 14,21 & 5,47  & 0,25  & 0,05 & 2.6***  & -0,00  & -0,00  & 954  & 1 & 5  & HV & 3  & 20,66 & 6,09  & 1,68  & 0,05 & 3.39*** & -0,00  & -0,03 & 867 \\
0 & 5  & HV & 5  & 9,11  & 3,15  & 0,08  & 0,05 & 2.89*** & -0,00  & -0,00  & 906  & 1 & 5  & HV & 5  & 14,2  & 3,82  & 1,67  & 0,05 & 3.71*** & -0,00  & -0,02 & 831 \\
0 & 5  & HV & 10 & 8,52  & 3,3   & 0,1   & 0,21 & 2.58*** & -0,00  & -0,00  & 860  & 1 & 5  & HV & 10 & 13,25 & 3,85  & 1,67  & 0,07 & 3.44*** & -0,00  & -0,01 & 799 \\
0 & 5  & HV & 21 & 11,46 & 4,74  & 0,07  & 0,02 & 2.42*** & -0,00  & -0,00  & 828  & 1 & 5  & HV & 21 & 16,99 & 5,43  & 1,59  & 0,07 & 3.13*** & -0,00  & -0,01 & 762 \\
0 & 5  & HV & 63 & 22,96 & 11,52 & 0,07  & 0,01 & 1.99*** & -0,00  & -0,00  & 764  & 1 & 5  & HV & 63 & 30,33 & 12,28 & 1,14  & 0,05 & 2.47*** & -0,00  & -0,01 & 746 \\
\arrayrulecolor{black!10}\midrule
0 & 5  & YZ & 3  & 5,33  & 2,69  & 0,14  & 0,34 & 1.98*** & -0,00  & -0,00  & 773  & 1 & 5  & YZ & 3  & 8,33  & 3,46  & 1,68  & 0,09 & 2.41*** & -0,00  & -0,01 & 717 \\
0 & 5  & YZ & 5  & 6,65  & 3,06  & 0,2   & 0,34 & 2.18*** & -0,00  & -0,00  & 823  & 1 & 5  & YZ & 5  & 10,79 & 3,68  & 1,67  & 0,07 & 2.93*** & -0,00  & -0,01 & 762 \\
0 & 5  & YZ & 10 & 9,19  & 3,78  & 0,1   & 0,02 & 2.43*** & -0,00  & -0,00  & 856  & 1 & 5  & YZ & 10 & 14,21 & 4,43  & 1,67  & 0,07 & 3.2***  & -0,00  & -0,01 & 797 \\
0 & 5  & YZ & 21 & 16,16 & 7,72  & 0,2   & 0,02 & 2.09*** & -0,00  & -0,00  & 862  & 1 & 5  & YZ & 21 & 23,16 & 8,39  & 1,67  & 0,05 & 2.76*** & -0,00  & -0,01 & 801 \\
0 & 5  & YZ & 63 & 26,47 & 12,49 & 0,07  & 0,01 & 2.12*** & -0,00  & -0,00  & 861  & 1 & 5  & YZ & 63 & 36,52 & 13,36 & 1,68  & 0,05 & 2.73*** & -0,00  & -0,02 & 827 \\
0 & 10 & GK & 3  & 7,99  & 3,17  & 0,01  & 0,01 & 2.52*** & 0,01  & -0,00  & 1101 & 1 & 10 & GK & 3  & 9,27  & 3,33  & 0,01  & 0,01 & 2.79*** & -0,00  & -0,00  & 987 \\
\arrayrulecolor{black!50}\midrule
0 & 10 & GK & 5  & 8,88  & 4,11  & 0,01  & 0,01 & 2.16*** & 0,01  & -0,00  & 1101 & 1 & 10 & GK & 5  & 10,06 & 4,13  & 0,01  & 0,01 & 2.43*** & -0,00  & -0,00  & 989 \\
0 & 10 & GK & 10 & 10,16 & 5,16  & 0,01  & 0,01 & 1.97*** & 0,01  & 0,01  & 1100 & 1 & 10 & GK & 10 & 11,59 & 5,41  & 0,01  & 0,01 & 2.14*** & -0,00  & -0,00  & 989 \\
0 & 10 & GK & 21 & 11,12 & 5,6   & 0,01  & 0,01 & 1.98*** & 0,01  & -0,00  & 1096 & 1 & 10 & GK & 21 & 12,36 & 5,82  & 0,01  & 0,01 & 2.13*** & -0,00  & -0,00  & 985 \\
0 & 10 & GK & 63 & 11,8  & 5,68  & 0,01  & 0,01 & 2.08*** & 0,01  & -0,00  & 1094 & 1 & 10 & GK & 63 & 13,7  & 6     & 0,01  & 0,01 & 2.28*** & -0,00  & -0,00  & 979 \\
\arrayrulecolor{black!10}\midrule
0 & 10 & HV & 3  & 6,05  & 1,62  & 0,01  & 0,01 & 3.74*** & -0,00  & -0,00  & 1075 & 1 & 10 & HV & 3  & 6,65  & 1,6   & 0,01  & 0,01 & 4.15*** & -0,00  & -0,00  & 960 \\
0 & 10 & HV & 5  & 5,36  & 1,35  & 0,01  & 0,01 & 3.97*** & -0,00  & -0,00  & 1065 & 1 & 10 & HV & 5  & 5,86  & 1,43  & 0,01  & 0,01 & 4.11*** & -0,00  & -0,00  & 950 \\
0 & 10 & HV & 10 & 5,4   & 1,41  & 0,01  & 0,01 & 3.84*** & -0,00  & -0,00  & 1068 & 1 & 10 & HV & 10 & 6,06  & 1,51  & 0,01  & 0,01 & 4.01*** & -0,00  & -0,00  & 950 \\
0 & 10 & HV & 21 & 6,42  & 1,94  & 0,01  & 0,01 & 3.31*** & -0,00  & -0,00  & 1084 & 1 & 10 & HV & 21 & 7,37  & 2,16  & 0,01  & 0,01 & 3.41*** & -0,00  & -0,00  & 969 \\
0 & 10 & HV & 63 & 10,29 & 5,45  & 0,01  & 0,01 & 1.89*** & -0,00  & -0,00  & 1066 & 1 & 10 & HV & 63 & 11,65 & 5,73  & 0,01  & 0,01 & 2.03*** & -0,00  & -0,00  & 954 \\
\arrayrulecolor{black!10}\midrule
0 & 10 & YZ & 3  & 4,61  & 1,32  & 0,01  & 0,01 & 3.5***  & -0,00  & -0,00  & 1036 & 1 & 10 & YZ & 3  & 5,07  & 1,41  & 0,01  & 0,01 & 3.6***  & -0,00  & -0,00  & 922 \\
0 & 10 & YZ & 5  & 5,02  & 1,34  & 0,01  & 0,01 & 3.74*** & -0,00  & -0,00  & 1065 & 1 & 10 & YZ & 5  & 5,66  & 1,44  & 0,01  & 0,01 & 3.93*** & -0,00  & -0,00  & 944 \\
0 & 10 & YZ & 10 & 5,45  & 1,42  & 0,01  & 0,01 & 3.83*** & -0,00  & -0,00  & 1079 & 1 & 10 & YZ & 10 & 6,32  & 1,54  & 0,01  & 0,01 & 4.1***  & -0,00  & -0,00  & 964 \\
0 & 10 & YZ & 21 & 8,22  & 3,58  & 0,01  & 0,01 & 2.3***  & -0,00  & -0,00  & 1090 & 1 & 10 & YZ & 21 & 9,36  & 3,71  & 0,01  & 0,01 & 2.52*** & -0,00  & -0,00  & 976 \\
0 & 10 & YZ & 63 & 10,9  & 5,56  & 0,01  & 0,01 & 1.96*** & -0,00  & -0,00  & 1070 & 1 & 10 & YZ & 63 & 12,61 & 5,94  & 0,01  & 0,01 & 2.12*** & -0,00  & -0,00  & 960 \\ \arrayrulecolor{black}\bottomrule
\end{tabular}
}
\source{Column N contains the total number of positions opened by the specified strategy. Moneyness (\%OTM) is expressed as a percentage of out-of-the-money. *, **, and *** denote statistical significance at the 10\%, 5\%, and 1\% levels, respectively, from the probabilistic Sharpe Ratio test against the benchmark buy-and-hold (B\&H) strategy. Volatility estimators abbreviations: HV, GK, YZ are historical volatility, Garman-Klass estimator, and Yang-Zhang estimator respectively. The estimator memory values are expressed in number of days used to estimate the volatility.}
\end{table}
\begin{table}[!ht]
\caption{Results of Short Put Strategies with Kelly Criterion Sizing with 3 or 5 Days to Expiration}
\label{tab:kelly-results-3-5}
\resizebox{\columnwidth}{!}{%
\begin{tabular}{@{}cccccccccccc|cccccccccccc@{}}
\toprule
\textbf{DTE} &
  \textbf{\%OTM} &
  \textbf{EST.} &
  \makecell{\textbf{EST.}\\\textbf{MEM.}} &
  \makecell{\textbf{aRC}\\\textbf{(\%)}} &
  \makecell{\textbf{aSD}\\\textbf{(\%)}} &
  \makecell{\textbf{MD}\\\textbf{(\%)}} &
  \textbf{MLD} &
  \makecell{\textbf{IR}\\\textbf{(\%)}} &
  \makecell{\textbf{VaR}\\\textbf{(\%)}} &
  \makecell{\textbf{CVaR}\\\textbf{(\%)}} &
  \textbf{N} &
  \textbf{DTE} &
  \textbf{\%OTM} &
  \textbf{EST.} &
  \makecell{\textbf{EST.}\\\textbf{MEM.}} &
  \makecell{\textbf{aRC}\\\textbf{(\%)}} &
  \makecell{\textbf{aSD}\\\textbf{(\%)}} &
  \makecell{\textbf{MD}\\\textbf{(\%)}} &
  \textbf{MLD} &
  \textbf{IR} &
  \makecell{\textbf{VaR}\\\textbf{(\%)}} &
  \makecell{\textbf{CVaR}\\\textbf{(\%)}} &
  \textbf{N} \\ \midrule
3 & 0  & GK & 3  & 18,16 & 64,02 & 75,5  & 1,26 & 0.28    & -5,25 & -12,28 & 716 & 5 & 0  & GK & 3  & 19,23  & 69,58 & 78,12 & 1,5  & 0.28   & -6,95 & -13    & 882 \\
3 & 0  & GK & 5  & 26,02 & 60,85 & 64,52 & 1,24 & 0.43    & -5,23 & -11,4  & 707 & 5 & 0  & GK & 5  & 14,48  & 69,19 & 75,85 & 1,54 & 0.21   & -7,22 & -12,85 & 876 \\
3 & 0  & GK & 10 & 33,57 & 58,26 & 64,98 & 1,07 & 0.58    & -5,21 & -10,68 & 669 & 5 & 0  & GK & 10 & 17,72  & 68,49 & 78,22 & 1,17 & 0.26   & -6,97 & -12,62 & 872 \\
3 & 0  & GK & 21 & 11,43 & 63,83 & 78,06 & 1,86 & 0.18    & -4,73 & -11,75 & 615 & 5 & 0  & GK & 21 & -5,23  & 70,07 & 81,63 & 2,63 & -0.07  & -7,06 & -13,29 & 844 \\
3 & 0  & GK & 63 & 9,43  & 65,62 & 81,13 & 1,95 & 0.14    & -4,79 & -11,98 & 529 & 5 & 0  & GK & 63 & -5,33  & 69,01 & 82,03 & 3,47 & -0.08  & -6,88 & -12,82 & 782 \\
\arrayrulecolor{black!10}\midrule
3 & 0  & HV & 3  & 23,21 & 61,93 & 66,14 & 1,71 & 0.37    & -6,26 & -11,98 & 649 & 5 & 0  & HV & 3  & 42,02  & 60,77 & 70,19 & 1,31 & 0.69   & -7,23 & -11,43 & 856 \\
3 & 0  & HV & 5  & 46,51 & 45,48 & 39,52 & 1,4  & 1.02    & -4,26 & -8,28  & 573 & 5 & 0  & HV & 5  & 34,88  & 55,29 & 60,26 & 1,21 & 0.63   & -5,75 & -10,37 & 797 \\
3 & 0  & HV & 10 & 43,83 & 39,64 & 30,86 & 0,87 & 1.11    & -3,36 & -6,92  & 482 & 5 & 0  & HV & 10 & 7,02   & 52,22 & 61,56 & 1,76 & 0.13   & -5,04 & -9,92  & 734 \\
3 & 0  & HV & 21 & 45,28 & 38,64 & 32,91 & 1,52 & 1.17    & -2,32 & -6,22  & 410 & 5 & 0  & HV & 21 & -7,48  & 51,43 & 65,73 & 2,15 & -0.15  & -5,16 & -9,91  & 667 \\
3 & 0  & HV & 63 & 0,19  & 54,56 & 79,86 & 3,52 & 0.0     & -1,93 & -8,74  & 318 & 5 & 0  & HV & 63 & -24,73 & 60,07 & 86,13 & 3,85 & -0.41  & -4,63 & -11,15 & 572 \\
\arrayrulecolor{black!10}\midrule
3 & 0  & YZ & 3  & 24,71 & 29,69 & 40,96 & 0,66 & 0.83    & -1,2  & -4,6   & 328 & 5 & 0  & YZ & 3  & 17,6   & 41,14 & 68,67 & 1,29 & 0.43   & -3,75 & -7,77  & 635 \\
3 & 0  & YZ & 5  & 36,11 & 30,97 & 35,62 & 0,81 & 1.17    & -1,72 & -4,75  & 394 & 5 & 0  & YZ & 5  & 9,33   & 47,58 & 70,89 & 1,5  & 0.2    & -4,5  & -9,15  & 686 \\
3 & 0  & YZ & 10 & 41,57 & 36,22 & 36,54 & 1,13 & 1.15    & -2,18 & -5,8   & 416 & 5 & 0  & YZ & 10 & -0,55  & 51,37 & 71,83 & 1,69 & -0.01  & -4,93 & -9,96  & 697 \\
3 & 0  & YZ & 21 & 20,04 & 44,11 & 49,42 & 2,05 & 0.45    & -2,61 & -7,71  & 411 & 5 & 0  & YZ & 21 & -13,11 & 54,57 & 75    & 3,85 & -0.24  & -5,8  & -10,94 & 681 \\
3 & 0  & YZ & 63 & 6,48  & 57,42 & 79,27 & 2,05 & 0.11    & -2,74 & -9,47  & 354 & 5 & 0  & YZ & 63 & -18,23 & 62,41 & 83,37 & 3,85 & -0.29  & -5,12 & -11,62 & 628 \\
\arrayrulecolor{black!50}\midrule
3 & 2  & GK & 3  & 2,05  & 52,03 & 64,97 & 1,22 & 0.04    & -3,28 & -10,33 & 827 & 5 & 2  & GK & 3  & 11,19  & 56,99 & 82,5  & 1,29 & 0.2    & -4,51 & -10,48 & 948 \\
3 & 2  & GK & 5  & 6,81  & 50,98 & 64,99 & 1,19 & 0.13    & -3,5  & -9,94  & 825 & 5 & 2  & GK & 5  & 13,61  & 56,33 & 81,58 & 1,21 & 0.24   & -4,63 & -10,44 & 953 \\
3 & 2  & GK & 10 & 8,83  & 50    & 71,66 & 1,86 & 0.18    & -3,55 & -9,76  & 813 & 5 & 2  & GK & 10 & 18,83  & 56,69 & 82,3  & 0,98 & 0.33   & -4,48 & -10,41 & 947 \\
3 & 2  & GK & 21 & 8,36  & 53,72 & 79,32 & 2,28 & 0.16    & -2,98 & -9,77  & 796 & 5 & 2  & GK & 21 & 15,49  & 57,53 & 83,89 & 1,24 & 0.27   & -4,63 & -10,49 & 936 \\
3 & 2  & GK & 63 & 7,99  & 55,2  & 80,17 & 2,28 & 0.14    & -2,99 & -9,97  & 741 & 5 & 2  & GK & 63 & 14,26  & 57,97 & 84,03 & 1,24 & 0.25   & -4,5  & -10,61 & 904 \\
\arrayrulecolor{black!10}\midrule
3 & 2  & HV & 3  & 8,37  & 46,9  & 59,77 & 1,32 & 0.18    & -3,55 & -9,34  & 767 & 5 & 2  & HV & 3  & 43,51  & 44,66 & 58,6  & 1,29 & 0.97   & -4,42 & -8,83  & 914 \\
3 & 2  & HV & 5  & 37,46 & 33,56 & 27,01 & 0,48 & 1.12    & -1,78 & -6,37  & 731 & 5 & 2  & HV & 5  & 43,09  & 41,91 & 42,87 & 0,76 & 1.03   & -3,89 & -8,07  & 892 \\
3 & 2  & HV & 10 & 35,89 & 31,73 & 25,43 & 0,43 & 1.13    & -1,38 & -5,62  & 668 & 5 & 2  & HV & 10 & 16,86  & 42,93 & 55,76 & 1,25 & 0.39   & -3,23 & -8,12  & 857 \\
3 & 2  & HV & 21 & 31,44 & 33,28 & 40,86 & 0,85 & 0.94    & -1,1  & -5,77  & 610 & 5 & 2  & HV & 21 & 5,4    & 44,77 & 65,92 & 1,29 & 0.12   & -3,61 & -8,64  & 833 \\
3 & 2  & HV & 63 & 4,64  & 47,95 & 78,24 & 2,28 & 0.1     & -1,25 & -7,36  & 530 & 5 & 2  & HV & 63 & -4,61  & 52,84 & 83,75 & 2,63 & -0.09  & -2,96 & -9,3   & 772 \\
\arrayrulecolor{black!10}\midrule
3 & 2  & YZ & 3  & 21,43 & 24,74 & 34,82 & 1,17 & 0.87    & -0,59 & -3,67  & 569 & 5 & 2  & YZ & 3  & 25,25  & 35,64 & 53,06 & 1,29 & 0.71   & -2,39 & -6,79  & 813 \\
3 & 2  & YZ & 5  & 33,68 & 26,32 & 33,12 & 1,17 & 1.28    & -1,13 & -4,11  & 631 & 5 & 2  & YZ & 5  & 28,31  & 39,56 & 52,19 & 1,5  & 0.72   & -2,98 & -7,31  & 852 \\
3 & 2  & YZ & 10 & 30,5  & 30,53 & 25,1  & 1,12 & 1.0     & -1,57 & -5,26  & 653 & 5 & 2  & YZ & 10 & 15,03  & 43,7  & 58,98 & 1,29 & 0.34   & -3,55 & -8,32  & 866 \\
3 & 2  & YZ & 21 & 7,45  & 42,95 & 67,55 & 1,94 & 0.17    & -1,37 & -7,3   & 635 & 5 & 2  & YZ & 21 & 1,29   & 48,71 & 75,49 & 2,63 & 0.03   & -3,59 & -9,39  & 847 \\
3 & 2  & YZ & 63 & 6,16  & 50,11 & 77,91 & 2,28 & 0.12    & -1,81 & -8,02  & 585 & 5 & 2  & YZ & 63 & 1,43   & 54,89 & 83,81 & 2,63 & 0.03   & -3,3  & -9,78  & 823 \\
\arrayrulecolor{black!50}\midrule
3 & 5  & GK & 3  & 22,25 & 32,47 & 50,56 & 0,49 & 0.69    & -0,45 & -4,15  & 878 & 5 & 5  & GK & 3  & 19,15  & 41,65 & 71,71 & 0,74 & 0.46   & -1,32 & -5,89  & 971 \\
3 & 5  & GK & 5  & 23,44 & 32,72 & 51,65 & 0,48 & 0.72    & -0,44 & -4,18  & 878 & 5 & 5  & GK & 5  & 20,75  & 41,66 & 72,16 & 0,76 & 0.5    & -1,46 & -5,9   & 974 \\
3 & 5  & GK & 10 & 20,52 & 34,65 & 61,74 & 0,89 & 0.59    & -0,36 & -4,25  & 875 & 5 & 5  & GK & 10 & 26,2   & 42,38 & 73,3  & 0,63 & 0.62   & -1,29 & -5,93  & 972 \\
3 & 5  & GK & 21 & 18,74 & 38,21 & 68,28 & 1,2  & 0.49    & -0,4  & -4,44  & 870 & 5 & 5  & GK & 21 & 24,25  & 42,53 & 73,39 & 0,77 & 0.57   & -1,29 & -5,88  & 968 \\
3 & 5  & GK & 63 & 25,85 & 38,5  & 68,35 & 0,92 & 0.67    & -0,35 & -4,3   & 861 & 5 & 5  & GK & 63 & 28,48  & 42,62 & 73,43 & 0,63 & 0.67   & -1,23 & -5,81  & 960 \\
\arrayrulecolor{black!10}\midrule
3 & 5  & HV & 3  & 23,29 & 26,8  & 33,09 & 0,53 & 0.87    & -0,32 & -3,29  & 834 & 5 & 5  & HV & 3  & 38,24  & 31,08 & 44,28 & 1,04 & 1.23   & -1,22 & -4,78  & 952 \\
3 & 5  & HV & 5  & 32,47 & 16,08 & 12,52 & 0,28 & 2.02*** & -0,2  & -2,08  & 821 & 5 & 5  & HV & 5  & 39,29  & 29,66 & 38,83 & 0,6  & 1.32   & -1,11 & -4,2   & 945 \\
3 & 5  & HV & 10 & 34,34 & 16,15 & 13,87 & 0,13 & 2.13*** & -0,15 & -1,78  & 802 & 5 & 5  & HV & 10 & 19,7   & 33    & 52,11 & 0,65 & 0.6    & -0,89 & -4,45  & 935 \\
3 & 5  & HV & 21 & 22,16 & 24,7  & 44,58 & 0,79 & 0.9     & -0,17 & -2,75  & 787 & 5 & 5  & HV & 21 & 12,69  & 35,52 & 63,85 & 1,21 & 0.36   & -0,92 & -4,91  & 918 \\
3 & 5  & HV & 63 & 13,6  & 36,87 & 68,37 & 1,6  & 0.37    & -0,15 & -3,59  & 750 & 5 & 5  & HV & 63 & 14,39  & 41,59 & 73,38 & 1,17 & 0.35   & -0,83 & -5,3   & 890 \\
\arrayrulecolor{black!10}\midrule
3 & 5  & YZ & 3  & 25,89 & 13,39 & 12,52 & 0,49 & 1.93*** & -0,11 & -1,25  & 753 & 5 & 5  & YZ & 3  & 39,55  & 23,65 & 20,74 & 0,94 & 1.67** & -0,74 & -3,21  & 909 \\
3 & 5  & YZ & 5  & 30,28 & 14,71 & 12,51 & 0,21 & 2.06*** & -0,15 & -1,6   & 795 & 5 & 5  & YZ & 5  & 39,09  & 26,25 & 27,4  & 0,65 & 1.49*  & -0,89 & -3,64  & 931 \\
3 & 5  & YZ & 10 & 28,7  & 18,32 & 20,91 & 0,27 & 1.57*   & -0,17 & -2,25  & 819 & 5 & 5  & YZ & 10 & 18,03  & 35,48 & 59,06 & 0,75 & 0.51   & -0,93 & -4,73  & 943 \\
3 & 5  & YZ & 21 & 12,3  & 33,88 & 63,41 & 1,67 & 0.36    & -0,18 & -3,56  & 806 & 5 & 5  & YZ & 21 & 8,28   & 40,56 & 73,26 & 2,38 & 0.2    & -0,93 & -5,52  & 940 \\
3 & 5  & YZ & 63 & 17,87 & 37,81 & 68,41 & 1,3  & 0.47    & -0,17 & -3,88  & 786 & 5 & 5  & YZ & 63 & 17,52  & 42,1  & 73,45 & 1,05 & 0.42   & -0,93 & -5,54  & 920 \\
\arrayrulecolor{black!50}\midrule
3 & 10 & GK & 3  & 27,54 & 10,87 & 4,98  & 0,04 & 2.53*** & -0,01 & -0,47  & 894 & 5 & 10 & GK & 3  & 16,84  & 22,85 & 42,97 & 0,25 & 0.74   & -0,11 & -2,02  & 979 \\
3 & 10 & GK & 5  & 21,94 & 14,95 & 19,57 & 0,04 & 1.47*   & -0,01 & -0,8   & 894 & 5 & 10 & GK & 5  & 16,37  & 22,96 & 43,35 & 0,3  & 0.71   & -0,11 & -2,04  & 979 \\
3 & 10 & GK & 10 & 20,44 & 15,66 & 19,69 & 0,04 & 1.3     & -0,01 & -0,89  & 895 & 5 & 10 & GK & 10 & 15,89  & 22,97 & 43,77 & 0,36 & 0.69   & -0,11 & -2,05  & 979 \\
3 & 10 & GK & 21 & 22,25 & 15,88 & 19,57 & 0,04 & 1.4     & -0,01 & -0,88  & 891 & 5 & 10 & GK & 21 & 15,78  & 23,09 & 43,87 & 0,37 & 0.68   & -0,12 & -2,05  & 977 \\
3 & 10 & GK & 63 & 23,5  & 16,05 & 19,66 & 0,04 & 1.46*   & -0,01 & -0,88  & 890 & 5 & 10 & GK & 63 & 17,21  & 23,14 & 43,87 & 0,29 & 0.74   & -0,1  & -2,05  & 973 \\
\arrayrulecolor{black!10}\midrule
3 & 10 & HV & 3  & 15,52 & 12,1  & 16,95 & 0,18 & 1.28    & 0     & -0,66  & 872 & 5 & 10 & HV & 3  & 16,32  & 18,86 & 33,1  & 0,25 & 0.87   & -0,1  & -1,72  & 968 \\
3 & 10 & HV & 5  & 17,87 & 7,01  & 4,96  & 0,08 & 2.55*** & 0     & -0,3   & 874 & 5 & 10 & HV & 5  & 16,16  & 20,05 & 31,99 & 0,34 & 0.81   & -0,1  & -1,57  & 967 \\
3 & 10 & HV & 10 & 16,98 & 7,7   & 4,96  & 0,02 & 2.21*** & -0,01 & -0,37  & 877 & 5 & 10 & HV & 10 & 9,61   & 19,45 & 33,78 & 1    & 0.49   & -0,1  & -1,65  & 969 \\
3 & 10 & HV & 21 & 16,81 & 12,54 & 15,57 & 0,04 & 1.34    & 0     & -0,71  & 872 & 5 & 10 & HV & 21 & 9,17   & 20,97 & 40,1  & 1,08 & 0.44   & -0,09 & -1,92  & 967 \\
3 & 10 & HV & 63 & 19,33 & 15,85 & 19,67 & 0,04 & 1.22    & 0     & -0,86  & 861 & 5 & 10 & HV & 63 & 12,61  & 23,02 & 43,89 & 0,57 & 0.55   & -0,1  & -2,02  & 955 \\
\arrayrulecolor{black!10}\midrule
3 & 10 & YZ & 3  & 16,18 & 6,94  & 4,96  & 0,02 & 2.33*** & 0     & -0,21  & 860 & 5 & 10 & YZ & 3  & 20,08  & 15,55 & 20,16 & 0,37 & 1.29   & -0,09 & -1,11  & 959 \\
3 & 10 & YZ & 5  & 16,34 & 7,46  & 4,96  & 0,04 & 2.19*** & -0,01 & -0,33  & 872 & 5 & 10 & YZ & 5  & 18,05  & 17,03 & 22,74 & 0,33 & 1.06   & -0,1  & -1,3   & 971 \\
3 & 10 & YZ & 10 & 17,03 & 9,18  & 8,71  & 0,04 & 1.86*** & -0,01 & -0,53  & 880 & 5 & 10 & YZ & 10 & 11,5   & 20,07 & 36,34 & 0,63 & 0.57   & -0,1  & -1,77  & 972 \\
3 & 10 & YZ & 21 & 17,79 & 15,55 & 19,61 & 0,04 & 1.14    & 0     & -0,87  & 883 & 5 & 10 & YZ & 21 & 11,87  & 22,76 & 43,87 & 0,73 & 0.52   & -0,11 & -2,03  & 969 \\
3 & 10 & YZ & 63 & 19,92 & 15,72 & 19,69 & 0,04 & 1.27    & 0     & -0,87  & 870 & 5 & 10 & YZ & 63 & 14,47  & 23,04 & 43,77 & 0,43 & 0.63   & -0,1  & -2,04  & 964 \\ \arrayrulecolor{black}\bottomrule
\end{tabular}
}
\source{Column N contains the total number of positions opened by the specified strategy. Moneyness (\%OTM) is expressed as a percentage of out-of-the-money. *, **, and *** denote statistical significance at the 10\%, 5\%, and 1\% levels, respectively, from the probabilistic Sharpe Ratio test against the benchmark buy-and-hold (B\&H) strategy. Volatility estimators abbreviations: HV, GK, YZ are historical volatility, Garman-Klass estimator, and Yang-Zhang estimator respectively. The estimator memory values are expressed in number of days used to estimate the volatility.}
\end{table}

\clearpage

\section{Short Put Strategies with VIX Sizing}
\begin{table}[!ht]
\caption{Results of Short Put Strategies with VIX Sizing with 0 or 1 Days to Expiration}
\label{tab:vix-results-0-1}
\resizebox{\columnwidth}{!}{%
\begin{tabular}{@{}cccccccccccc|cccccccccccc@{}}
\toprule
\textbf{DTE} &
  \textbf{\%OTM} &
  \textbf{VIX} &
  \makecell{\textbf{VIX}\\\textbf{MEM.}} &
  \makecell{\textbf{aRC}\\\textbf{(\%)}} &
  \makecell{\textbf{aSD}\\\textbf{(\%)}} &
  \makecell{\textbf{MD}\\\textbf{(\%)}} &
  \textbf{MLD} &
  \makecell{\textbf{IR}\\\textbf{(\%)}} &
  \makecell{\textbf{VaR}\\\textbf{(\%)}} &
  \makecell{\textbf{CVaR}\\\textbf{(\%)}} &
  \textbf{N} &
  \textbf{DTE} &
  \textbf{\%OTM} &
  \textbf{VIX} &
  \makecell{\textbf{VIX}\\\textbf{MEM.}} &
  \makecell{\textbf{aRC}\\\textbf{(\%)}} &
  \makecell{\textbf{aSD}\\\textbf{(\%)}} &
  \makecell{\textbf{MD}\\\textbf{(\%)}} &
  \textbf{MLD} &
  \textbf{IR} &
  \makecell{\textbf{VaR}\\\textbf{(\%)}} &
  \makecell{\textbf{CVaR}\\\textbf{(\%)}} &
  \textbf{N} \\ \midrule
0 & 0 & VIX30D & 21 & 23,38 & 39,28 & 57,43 & 1,89 & 0.6 & -4,07 & -7,69 & 1004 & 1 & 0 & VIX30D & 21 & 34,81 & 45,06 & 61,55 & 1,53 & 0.77 & -4,47 & -8,91 & 900 \\
0 & 0 & VIX30D & 42 & 12,7 & 38,42 & 61,35 & 1,96 & 0.33 & -4,01 & -7,41 & 1053 & 1 & 0 & VIX30D & 42 & 26,39 & 43,02 & 63,2 & 1,6 & 0.61 & -4 & -8,25 & 943 \\
0 & 0 & VIX30D & 63 & 16,1 & 37,07 & 59,49 & 1,96 & 0.43 & -3,88 & -7,21 & 1072 & 1 & 0 & VIX30D & 63 & 32,98 & 40,82 & 60,76 & 1,6 & 0.81 & -3,87 & -7,85 & 959 \\
0 & 0 & VIX30D & 84 & 15,17 & 35,63 & 58,27 & 1,96 & 0.43 & -3,82 & -7 & 1080 & 1 & 0 & VIX30D & 84 & 29,97 & 39,43 & 61,08 & 1,6 & 0.76 & -3,81 & -7,6 & 967 \\
0 & 0 & VIX30D & 126 & 16,56 & 34,11 & 52,26 & 1,96 & 0.49 & -4 & -6,64 & 1084 & 1 & 0 & VIX30D & 126 & 32,72 & 37,75 & 54,59 & 1,6 & 0.87 & -4,11 & -7,21 & 974 \\
0 & 0 & VIX30D & 252 & 13,83 & 31 & 46,75 & 1,96 & 0.45 & -3,77 & -6,01 & 1092 & 1 & 0 & VIX30D & 252 & 30,41 & 34,2 & 49,14 & 1,58 & 0.89 & -3,95 & -6,5 & 980 \\
\arrayrulecolor{black!10}\midrule
0 & 0 & VIX9D & 21 & 35,29 & 37,38 & 54,25 & 1,96 & 0.94 & -3,42 & -7,24 & 1032 & 1 & 0 & VIX9D & 21 & 43,85 & 42,33 & 59,22 & 1,52 & 1.04 & -3,66 & -8,33 & 925 \\
0 & 0 & VIX9D & 42 & 19,76 & 36,99 & 57,57 & 1,96 & 0.53 & -3,58 & -7,02 & 1060 & 1 & 0 & VIX9D & 42 & 31,81 & 40,75 & 60,22 & 1,6 & 0.78 & -3,66 & -7,8 & 950 \\
0 & 0 & VIX9D & 63 & 20,66 & 36,03 & 56,51 & 1,96 & 0.57 & -3,58 & -6,86 & 1076 & 1 & 0 & VIX9D & 63 & 35,14 & 39,18 & 58,78 & 1,6 & 0.9 & -3,64 & -7,46 & 963 \\
0 & 0 & VIX9D & 84 & 19,78 & 34,69 & 53,46 & 1,96 & 0.57 & -3,51 & -6,72 & 1083 & 1 & 0 & VIX9D & 84 & 33,55 & 37,76 & 56,55 & 1,6 & 0.89 & -3,62 & -7,24 & 970 \\
0 & 0 & VIX9D & 126 & 19,25 & 32,81 & 46,67 & 1,96 & 0.59 & -3,59 & -6,3 & 1084 & 1 & 0 & VIX9D & 126 & 33,49 & 35,62 & 49,5 & 1,6 & 0.94 & -3,7 & -6,8 & 973 \\
0 & 0 & VIX9D & 252 & 14,2 & 29,58 & 41,84 & 1,96 & 0.48 & -3,48 & -5,7 & 1089 & 1 & 0 & VIX9D & 252 & 28,01 & 32,22 & 45,47 & 1,6 & 0.87 & -3,64 & -6,12 & 976 \\
\arrayrulecolor{black!50}\midrule
0 & 2 & VIX30D & 21 & 12,2 & 15,27 & 26,84 & 1,67 & 0.8 & -0,00 & -0,74 & 1004 & 1 & 2 & VIX30D & 21 & 9,07 & 22,08 & 37,85 & 1,32 & 0.41 & -0,25 & -3,55 & 900 \\
0 & 2 & VIX30D & 42 & 4,35 & 15,87 & 30,84 & 1,86 & 0.27 & -0,00 & -1,18 & 1053 & 1 & 2 & VIX30D & 42 & 4,46 & 20,92 & 37,12 & 1,32 & 0.21 & -0,26 & -3,44 & 943 \\
0 & 2 & VIX30D & 63 & 3,47 & 15,69 & 33,07 & 1,86 & 0.22 & -0,00 & -1,42 & 1072 & 1 & 2 & VIX30D & 63 & 5,85 & 19,67 & 35,05 & 1,5 & 0.3 & -0,34 & -3,25 & 959 \\
0 & 2 & VIX30D & 84 & 3,73 & 14,26 & 30,35 & 1,86 & 0.26 & -0,00 & -1,5 & 1080 & 1 & 2 & VIX30D & 84 & 5,01 & 18,49 & 32,66 & 1,5 & 0.27 & -0,37 & -3,19 & 967 \\
0 & 2 & VIX30D & 126 & 5 & 12,72 & 26,53 & 1,86 & 0.39 & -0,00 & -1,32 & 1076 & 1 & 2 & VIX30D & 126 & 7,54 & 16,77 & 27,43 & 1,5 & 0.45 & -0,36 & -2,99 & 963 \\
0 & 2 & VIX30D & 252 & 5,78 & 9,83 & 20,52 & 1,86 & 0.59 & -0,00 & -1,35 & 1090 & 1 & 2 & VIX30D & 252 & 9,52 & 13,31 & 21,97 & 1,06 & 0.72 & -0,34 & -2,47 & 976 \\
\arrayrulecolor{black!10}\midrule
0 & 2 & VIX9D & 21 & 15,03 & 13,99 & 20,32 & 1,67 & 1.07 & -0,00 & -0,81 & 1032 & 1 & 2 & VIX9D & 21 & 12,98 & 20,19 & 33,56 & 1,32 & 0.64 & -0,27 & -3,19 & 925 \\
0 & 2 & VIX9D & 42 & 5,37 & 15,25 & 25,99 & 1,67 & 0.35 & -0,00 & -1,19 & 1060 & 1 & 2 & VIX9D & 42 & 5,86 & 19,44 & 34,45 & 1,32 & 0.3 & -0,31 & -3,21 & 950 \\
0 & 2 & VIX9D & 63 & 3,32 & 15,41 & 29,44 & 1,67 & 0.22 & -0,00 & -1,44 & 1076 & 1 & 2 & VIX9D & 63 & 5,3 & 18,8 & 33,54 & 1,32 & 0.28 & -0,34 & -3,09 & 963 \\
0 & 2 & VIX9D & 84 & 3,13 & 14,19 & 28,13 & 1,86 & 0.22 & -0,00 & -1,52 & 1083 & 1 & 2 & VIX9D & 84 & 4,79 & 17,58 & 30,88 & 1,32 & 0.27 & -0,39 & -3,01 & 970 \\
0 & 2 & VIX9D & 126 & 4,22 & 12,55 & 23,98 & 1,86 & 0.34 & -0,00 & -1,32 & 1080 & 1 & 2 & VIX9D & 126 & 6,09 & 15,75 & 25,35 & 1,5 & 0.39 & -0,44 & -2,78 & 967 \\
0 & 2 & VIX9D & 252 & 4,5 & 9,61 & 18,59 & 1,86 & 0.47 & -0,00 & -1,3 & 1091 & 1 & 2 & VIX9D & 252 & 7,38 & 12,29 & 19,88 & 1,16 & 0.6 & -0,32 & -2,28 & 977 \\
\arrayrulecolor{black!50}\midrule
0 & 5 & VIX30D & 21 & 7,74 & 2,05 & 0,42 & 0,1 & 3.77*** & -0,00 & -0,00 & 1004 & 1 & 5 & VIX30D & 21 & 11,04 & 2,59 & 0,96 & 0,09 & 4.26*** & -0,00 & -0,03 & 900 \\
0 & 5 & VIX30D & 42 & 6,21 & 1,14 & 0,47 & 0,1 & 5.44*** & -0,00 & -0,01 & 1053 & 1 & 5 & VIX30D & 42 & 9,1 & 1,6 & 0,84 & 0,12 & 5.69*** & -0,00 & -0,04 & 943 \\
0 & 5 & VIX30D & 63 & 5,77 & 0,86 & 0,48 & 0,07 & 6.69*** & -0,00 & -0,01 & 1072 & 1 & 5 & VIX30D & 63 & 8,51 & 1,34 & 1,12 & 0,12 & 6.33*** & -0,00 & -0,06 & 959 \\
0 & 5 & VIX30D & 84 & 5,47 & 0,71 & 0,4 & 0,07 & 7.67*** & -0,00 & -0,01 & 1080 & 1 & 5 & VIX30D & 84 & 8,17 & 1,24 & 1,25 & 0,12 & 6.6*** & -0,00 & -0,06 & 967 \\
0 & 5 & VIX30D & 126 & 5,14 & 0,58 & 0,32 & 0,1 & 8.79*** & -0,00 & -0,01 & 1088 & 1 & 5 & VIX30D & 126 & 8,08 & 1,2 & 1,13 & 0,08 & 6.76*** & -0,00 & -0,05 & 974 \\
0 & 5 & VIX30D & 252 & 4,54 & 0,48 & 0,21 & 0,13 & 9.48*** & -0,00 & -0,00 & 1094 & 1 & 5 & VIX30D & 252 & 7,24 & 0,97 & 0,86 & 0,06 & 7.46*** & -0,00 & -0,04 & 980 \\
\arrayrulecolor{black!10}\midrule
0 & 5 & VIX9D & 21 & 7,82 & 2,19 & 0,35 & 0,07 & 3.57*** & -0,00 & -0,00 & 1032 & 1 & 5 & VIX9D & 21 & 11,44 & 2,73 & 0,91 & 0,05 & 4.19*** & -0,00 & -0,03 & 925 \\
0 & 5 & VIX9D & 42 & 6,29 & 1,23 & 0,43 & 0,1 & 5.11*** & -0,00 & -0,01 & 1060 & 1 & 5 & VIX9D & 42 & 9,19 & 1,6 & 0,69 & 0,07 & 5.75*** & -0,00 & -0,04 & 950 \\
0 & 5 & VIX9D & 63 & 5,71 & 0,91 & 0,46 & 0,1 & 6.3*** & -0,00 & -0,01 & 1076 & 1 & 5 & VIX9D & 63 & 8,26 & 1,26 & 0,79 & 0,11 & 6.55*** & -0,00 & -0,05 & 963 \\
0 & 5 & VIX9D & 84 & 5,36 & 0,73 & 0,39 & 0,08 & 7.3*** & -0,00 & -0,01 & 1083 & 1 & 5 & VIX9D & 84 & 7,8 & 1,12 & 0,99 & 0,08 & 6.94*** & -0,00 & -0,05 & 970 \\
0 & 5 & VIX9D & 126 & 4,91 & 0,57 & 0,31 & 0,1 & 8.64*** & -0,00 & -0,01 & 1086 & 1 & 5 & VIX9D & 126 & 7,42 & 0,95 & 0,9 & 0,07 & 7.8*** & -0,00 & -0,04 & 973 \\
0 & 5 & VIX9D & 252 & 4,26 & 0,43 & 0,2 & 0,16 & 9.95*** & -0,00 & -0,00 & 1092 & 1 & 5 & VIX9D & 252 & 6,49 & 0,69 & 0,65 & 0,05 & 9.37*** & -0,00 & -0,03 & 978 \\
\arrayrulecolor{black!50}\midrule
0 & 10 & VIX30D & 21 & 2,64 & 0,39 & 0,01 & 0,02 & 6.74*** & -0,00 & -0,00 & 1004 & 1 & 10 & VIX30D & 21 & 2,9 & 0,47 & 0,01 & 0,02 & 6.15*** & -0,00 & -0,00 & 900 \\
0 & 10 & VIX30D & 42 & 2,42 & 0,22 & 0,01 & 0,01 & 11.21*** & -0,00 & -0,00 & 1053 & 1 & 10 & VIX30D & 42 & 2,6 & 0,26 & 0,01 & 0,01 & 9.98*** & -0,00 & -0,00 & 943 \\
0 & 10 & VIX30D & 63 & 2,38 & 0,17 & 0,01 & 0,01 & 13.86*** & -0,00 & -0,00 & 1072 & 1 & 10 & VIX30D & 63 & 2,53 & 0,21 & 0,01 & 0,01 & 12.24*** & -0,00 & -0,00 & 959 \\
0 & 10 & VIX30D & 84 & 2,33 & 0,15 & 0,01 & 0,01 & 15.67*** & -0,00 & -0,00 & 1080 & 1 & 10 & VIX30D & 84 & 2,46 & 0,18 & 0,01 & 0,01 & 13.65*** & -0,00 & -0,00 & 967 \\
0 & 10 & VIX30D & 126 & 2,26 & 0,13 & 0,01 & 0,01 & 17.79*** & -0,00 & -0,00 & 1080 & 1 & 10 & VIX30D & 126 & 2,4 & 0,17 & 0,01 & 0,01 & 13.86*** & -0,00 & -0,00 & 967 \\
0 & 10 & VIX30D & 252 & 2,07 & 0,11 & 0,01 & 0,01 & 18.15*** & -0,00 & -0,00 & 1092 & 1 & 10 & VIX30D & 252 & 2,17 & 0,16 & 0,01 & 0,01 & 13.87*** & -0,00 & -0,00 & 978 \\
\arrayrulecolor{black!10}\midrule
0 & 10 & VIX9D & 21 & 2,61 & 0,41 & -0,00 & 0,02 & 6.31*** & -0,00 & -0,00 & 1032 & 1 & 10 & VIX9D & 21 & 2,93 & 0,52 & 0,01 & 0,02 & 5.62*** & -0,00 & -0,00 & 925 \\
0 & 10 & VIX9D & 42 & 2,4 & 0,22 & 0,01 & 0,01 & 10.68*** & -0,00 & -0,00 & 1060 & 1 & 10 & VIX9D & 42 & 2,57 & 0,28 & 0,01 & 0,01 & 9.23*** & -0,00 & -0,00 & 950 \\
0 & 10 & VIX9D & 63 & 2,34 & 0,17 & 0,01 & 0,01 & 13.55*** & -0,00 & -0,00 & 1076 & 1 & 10 & VIX9D & 63 & 2,46 & 0,21 & 0,01 & 0,01 & 11.69*** & -0,00 & -0,00 & 963 \\
0 & 10 & VIX9D & 84 & 2,28 & 0,15 & 0,01 & 0,01 & 15.51*** & -0,00 & -0,00 & 1083 & 1 & 10 & VIX9D & 84 & 2,38 & 0,18 & 0,01 & 0,01 & 13.37*** & -0,00 & -0,00 & 970 \\
0 & 10 & VIX9D & 126 & 2,19 & 0,12 & 0,01 & 0,01 & 18.01*** & -0,00 & -0,00 & 1083 & 1 & 10 & VIX9D & 126 & 2,27 & 0,15 & 0,01 & 0,01 & 15.11*** & -0,00 & -0,00 & 970 \\
0 & 10 & VIX9D & 252 & 2 & 0,1 & 0,01 & 0,01 & 19.13*** & -0,00 & -0,00 & 1092 & 1 & 10 & VIX9D & 252 & 2,03 & 0,13 & 0,01 & 0,01 & 16.14*** & -0,00 & -0,00 & 978 \\ \arrayrulecolor{black}\bottomrule
\end{tabular}%
}
\source{Column N contains the total number of positions opened by the specified strategy. Moneyness (\%OTM) is expressed as a percentage of out-of-the-money. *, **, and *** denote statistical significance at the 10\%, 5\%, and 1\% levels, respectively, from the probabilistic Sharpe Ratio test against the benchmark buy-and-hold (B\&H) strategy. The VIX memory values are expressed in number of days used to calculate the percentile rank of the most recent VIX value.}
\end{table}
\begin{table}[!ht]
\caption{Results of Short Put Strategies with VIX Sizing with 3 or 5 Days to Expiration}
\label{tab:vix-results-3-5}
\resizebox{\columnwidth}{!}{%
\begin{tabular}{@{}cccccccccccc|cccccccccccc@{}}
\toprule
\textbf{DTE} &
  \textbf{\%OTM} &
  \textbf{VIX} &
  \makecell{\textbf{VIX}\\\textbf{MEM.}} &
  \makecell{\textbf{aRC}\\\textbf{(\%)}} &
  \makecell{\textbf{aSD}\\\textbf{(\%)}} &
  \makecell{\textbf{MD}\\\textbf{(\%)}} &
  \textbf{MLD} &
  \makecell{\textbf{IR}\\\textbf{(\%)}} &
  \makecell{\textbf{VaR}\\\textbf{(\%)}} &
  \makecell{\textbf{CVaR}\\\textbf{(\%)}} &
  \textbf{N} &
  \textbf{DTE} &
  \textbf{\%OTM} &
  \textbf{VIX} &
  \makecell{\textbf{VIX}\\\textbf{MEM.}} &
  \makecell{\textbf{aRC}\\\textbf{(\%)}} &
  \makecell{\textbf{aSD}\\\textbf{(\%)}} &
  \makecell{\textbf{MD}\\\textbf{(\%)}} &
  \textbf{MLD} &
  \textbf{IR} &
  \makecell{\textbf{VaR}\\\textbf{(\%)}} &
  \makecell{\textbf{CVaR}\\\textbf{(\%)}} &
  \textbf{N} \\ \midrule
3 & 0 & VIX30D & 21 & 21,24 & 55,31 & 68,66 & 1,2 & 0.38 & -4,93 & -10,91 & 812 & 5 & 0 & VIX30D & 21 & 51,19 & 42,86 & 48,58 & 1,46 & 1.19 & -4,52 & -8,11 & 894 \\
3 & 0 & VIX30D & 42 & 16,03 & 52,36 & 67,26 & 1,24 & 0.31 & -4,99 & -10,31 & 846 & 5 & 0 & VIX30D & 42 & 40,03 & 41,23 & 52,93 & 1,46 & 0.97 & -4,68 & -8,14 & 933 \\
3 & 0 & VIX30D & 63 & 23,94 & 49,26 & 67,84 & 1,24 & 0.49 & -4,53 & -9,8 & 862 & 5 & 0 & VIX30D & 63 & 43,4 & 40,12 & 52,58 & 1,54 & 1.08 & -4,61 & -7,88 & 952 \\
3 & 0 & VIX30D & 84 & 23,69 & 46,74 & 69,8 & 1,24 & 0.51 & -4,55 & -9,28 & 870 & 5 & 0 & VIX30D & 84 & 38,03 & 39,02 & 56,89 & 1,54 & 0.97 & -4,49 & -7,74 & 958 \\
3 & 0 & VIX30D & 126 & 30,7 & 43,58 & 62,78 & 1,24 & 0.7 & -4,87 & -8,57 & 875 & 5 & 0 & VIX30D & 126 & 42,05 & 37,27 & 51,3 & 1,54 & 1.13 & -4,41 & -7,26 & 964 \\
3 & 0 & VIX30D & 252 & 32,8 & 39,16 & 55,37 & 1,24 & 0.84 & -4,24 & -7,63 & 880 & 5 & 0 & VIX30D & 252 & 37,96 & 33,96 & 45,92 & 1,54 & 1.12 & -4,01 & -6,5 & 964 \\
\arrayrulecolor{black!10}\midrule
3 & 0 & VIX9D & 21 & 42,61 & 51,47 & 67,76 & 1,24 & 0.83 & -4,66 & -9,95 & 830 & 5 & 0 & VIX9D & 21 & 66,92 & 40,4 & 44,64 & 1,43 & 1.66** & -4,36 & -7,45 & 917 \\
3 & 0 & VIX9D & 42 & 31,12 & 49,04 & 65,54 & 1,24 & 0.63 & -4,55 & -9,52 & 854 & 5 & 0 & VIX9D & 42 & 47,98 & 39,04 & 47,95 & 1,54 & 1.23 & -4,31 & -7,5 & 940 \\
3 & 0 & VIX9D & 63 & 32,02 & 46,81 & 65,62 & 1,24 & 0.68 & -4,83 & -9,2 & 865 & 5 & 0 & VIX9D & 63 & 46,31 & 38,41 & 48,7 & 1,54 & 1.21 & -4,62 & -7,4 & 953 \\
3 & 0 & VIX9D & 84 & 31,06 & 44,27 & 66,52 & 1,24 & 0.7 & -4,28 & -8,79 & 870 & 5 & 0 & VIX9D & 84 & 40,68 & 37,35 & 50,43 & 1,54 & 1.09 & -4,28 & -7,34 & 956 \\
3 & 0 & VIX9D & 126 & 33,6 & 40,98 & 58,43 & 1,24 & 0.82 & -4,58 & -8,16 & 873 & 5 & 0 & VIX9D & 126 & 41,46 & 35,37 & 44,5 & 1,54 & 1.17 & -4,2 & -6,82 & 958 \\
3 & 0 & VIX9D & 252 & 31,04 & 37,04 & 51,8 & 1,24 & 0.84 & -4,1 & -7,25 & 876 & 5 & 0 & VIX9D & 252 & 35,81 & 32,18 & 40,92 & 1,54 & 1.11 & -3,96 & -6,12 & 962 \\
\arrayrulecolor{black!50}\midrule
3 & 2 & VIX30D & 21 & 1,05 & 36,46 & 58,05 & 1,19 & 0.03 & -1,67 & -6,81 & 812 & 5 & 2 & VIX30D & 21 & 44,06 & 25,38 & 28,78 & 1,32 & 1.74** & -1,9 & -4,7 & 894 \\
3 & 2 & VIX30D & 42 & -2,05 & 33,66 & 56,24 & 1,19 & -0.06 & -1,74 & -6,4 & 846 & 5 & 2 & VIX30D & 42 & 35,06 & 23,18 & 32,1 & 1,46 & 1.51* & -1,83 & -4,59 & 933 \\
3 & 2 & VIX30D & 63 & 1,89 & 30,72 & 55,35 & 1,19 & 0.06 & -1,53 & -5,95 & 862 & 5 & 2 & VIX30D & 63 & 34,57 & 22,21 & 33,31 & 1,46 & 1.56* & -1,9 & -4,53 & 952 \\
3 & 2 & VIX30D & 84 & 3,33 & 28,47 & 56,38 & 1,19 & 0.12 & -1,66 & -5,56 & 870 & 5 & 2 & VIX30D & 84 & 31,04 & 21,29 & 37 & 1,46 & 1.46* & -1,86 & -4,41 & 958 \\
3 & 2 & VIX30D & 126 & 8,85 & 24,78 & 48,83 & 1,19 & 0.36 & -1,8 & -5,02 & 873 & 5 & 2 & VIX30D & 126 & 32,69 & 19,73 & 33,04 & 1,46 & 1.66** & -1,95 & -4,07 & 964 \\
3 & 2 & VIX30D & 252 & 12,4 & 20,08 & 39,18 & 1,19 & 0.62 & -1,65 & -4,14 & 876 & 5 & 2 & VIX30D & 252 & 30,27 & 16,73 & 27,56 & 1,12 & 1.81** & -1,73 & -3,35 & 964 \\
\arrayrulecolor{black!10}\midrule
3 & 2 & VIX9D & 21 & 12,92 & 33,14 & 55,08 & 1,2 & 0.39 & -1,98 & -6,06 & 830 & 5 & 2 & VIX9D & 21 & 52,32 & 23,72 & 24,88 & 0,73 & 2.21*** & -1,81 & -4,27 & 917 \\
3 & 2 & VIX9D & 42 & 6,52 & 30,87 & 52,08 & 1,2 & 0.21 & -1,6 & -5,76 & 854 & 5 & 2 & VIX9D & 42 & 40,22 & 21,7 & 27,2 & 0,89 & 1.85** & -1,77 & -4,16 & 940 \\
3 & 2 & VIX9D & 63 & 5,98 & 28,79 & 52,22 & 1,2 & 0.21 & -1,7 & -5,54 & 865 & 5 & 2 & VIX9D & 63 & 36,1 & 21,01 & 27,72 & 1,29 & 1.72** & -1,89 & -4,21 & 953 \\
3 & 2 & VIX9D & 84 & 7,28 & 26,37 & 52,46 & 1,2 & 0.28 & -1,5 & -5,19 & 870 & 5 & 2 & VIX9D & 84 & 32,36 & 20,1 & 30,19 & 1,46 & 1.61** & -1,81 & -4,11 & 956 \\
3 & 2 & VIX9D & 126 & 10,28 & 22,72 & 44,41 & 1,24 & 0.45 & -1,6 & -4,67 & 872 & 5 & 2 & VIX9D & 126 & 31,89 & 18,36 & 26 & 1,43 & 1.74** & -1,78 & -3,81 & 958 \\
3 & 2 & VIX9D & 252 & 10,99 & 18,45 & 35,71 & 1,24 & 0.6 & -1,56 & -3,86 & 877 & 5 & 2 & VIX9D & 252 & 28,6 & 15,41 & 22,24 & 1,12 & 1.86** & -1,66 & -3,12 & 963 \\
\arrayrulecolor{black!50}\midrule
3 & 5 & VIX30D & 21 & 6,59 & 16,05 & 28 & 1 & 0.41 & -0,1 & -1,96 & 812 & 5 & 5 & VIX30D & 21 & 28,82 & 10,77 & 12,75 & 1,17 & 2.67*** & -0,35 & -1,55 & 894 \\
3 & 5 & VIX30D & 42 & 4,87 & 13,67 & 25,31 & 1 & 0.36 & -0,08 & -1,74 & 846 & 5 & 5 & VIX30D & 42 & 22,88 & 8,89 & 13,45 & 0,94 & 2.58*** & -0,31 & -1,46 & 933 \\
3 & 5 & VIX30D & 63 & 6,36 & 11,56 & 23,2 & 1 & 0.55 & -0,09 & -1,54 & 862 & 5 & 5 & VIX30D & 63 & 21,35 & 8,31 & 12,79 & 0,99 & 2.57*** & -0,3 & -1,45 & 952 \\
3 & 5 & VIX30D & 84 & 6,69 & 10,8 & 22,87 & 1 & 0.62 & -0,09 & -1,41 & 870 & 5 & 5 & VIX30D & 84 & 19,59 & 7,78 & 12,69 & 1,21 & 2.52*** & -0,34 & -1,4 & 958 \\
3 & 5 & VIX30D & 126 & 9,03 & 8,42 & 17,91 & 1 & 1.07 & -0,11 & -1,16 & 876 & 5 & 5 & VIX30D & 126 & 19,83 & 6,67 & 10,31 & 0,76 & 2.98*** & -0,32 & -1,25 & 964 \\
3 & 5 & VIX30D & 252 & 10,56 & 5,55 & 10,72 & 0,63 & 1.9** & -0,09 & -0,82 & 880 & 5 & 5 & VIX30D & 252 & 18,38 & 5,1 & 6,88 & 0,62 & 3.61*** & -0,31 & -0,93 & 964 \\
\arrayrulecolor{black!10}\midrule
3 & 5 & VIX9D & 21 & 10,56 & 14,77 & 25,38 & 1 & 0.71 & -0,11 & -1,73 & 830 & 5 & 5 & VIX9D & 21 & 31,04 & 10,29 & 10,73 & 0,6 & 3.02*** & -0,35 & -1,42 & 917 \\
3 & 5 & VIX9D & 42 & 7,63 & 12,57 & 22,91 & 1 & 0.61 & -0,1 & -1,56 & 854 & 5 & 5 & VIX9D & 42 & 24,34 & 8,48 & 12,04 & 0,73 & 2.87*** & -0,35 & -1,34 & 940 \\
3 & 5 & VIX9D & 63 & 7,25 & 10,92 & 21,9 & 1 & 0.66 & -0,1 & -1,43 & 865 & 5 & 5 & VIX9D & 63 & 21,21 & 7,92 & 12,04 & 0,86 & 2.68*** & -0,3 & -1,37 & 953 \\
3 & 5 & VIX9D & 84 & 7,51 & 9,88 & 20,97 & 1 & 0.76 & -0,1 & -1,3 & 870 & 5 & 5 & VIX9D & 84 & 19,32 & 7,38 & 11,52 & 0,92 & 2.62*** & -0,34 & -1,32 & 956 \\
3 & 5 & VIX9D & 126 & 8,75 & 7,63 & 16,52 & 1 & 1.15 & -0,1 & -1,06 & 873 & 5 & 5 & VIX9D & 126 & 18,63 & 6,13 & 8,89 & 0,75 & 3.04*** & -0,31 & -1,17 & 958 \\
3 & 5 & VIX9D & 252 & 9,5 & 4,86 & 9,67 & 0,65 & 1.96** & -0,09 & -0,75 & 878 & 5 & 5 & VIX9D & 252 & 17,01 & 4,52 & 5,5 & 0,6 & 3.77*** & -0,3 & -0,87 & 963 \\
\arrayrulecolor{black!50}\midrule
3 & 10 & VIX30D & 21 & 6,17 & 2,01 & 1,49 & 0,74 & 3.07*** & -0.00 & -0,11 & 812 & 5 & 10 & VIX30D & 21 & 11,16 & 3,08 & 1,66 & 0,08 & 3.63*** & -0,02 & -0,18 & 894 \\
3 & 10 & VIX30D & 42 & 5,11 & 1,23 & 1,07 & 0,62 & 4.15*** & -0.00 & -0,08 & 846 & 5 & 10 & VIX30D & 42 & 8,92 & 1,83 & 0,87 & 0,1 & 4.86*** & -0,03 & -0,15 & 933 \\
3 & 10 & VIX30D & 63 & 5 & 0,93 & 0,72 & 0,37 & 5.39*** & -0.00 & -0,07 & 862 & 5 & 10 & VIX30D & 63 & 8,37 & 1,51 & 1,01 & 0,1 & 5.54*** & -0,03 & -0,14 & 952 \\
3 & 10 & VIX30D & 84 & 4,8 & 0,77 & 0,53 & 0,25 & 6.25*** & -0.00 & -0,06 & 870 & 5 & 10 & VIX30D & 84 & 7,86 & 1,31 & 0,88 & 0,1 & 5.99*** & -0,03 & -0,14 & 958 \\
3 & 10 & VIX30D & 126 & 4,71 & 0,67 & 0,38 & 0,25 & 7.08*** & -0.00 & -0,05 & 873 & 5 & 10 & VIX30D & 126 & 7,43 & 1,13 & 0,58 & 0,1 & 6.58*** & -0,03 & -0,12 & 964 \\
3 & 10 & VIX30D & 252 & 4,31 & 0,6 & 0,41 & 0,25 & 7.18*** & -0.00 & -0,04 & 879 & 5 & 10 & VIX30D & 252 & 6,49 & 0,93 & 0,48 & 0,15 & 6.99*** & -0,02 & -0,1 & 964 \\
3 & 10 & VIX9D & 21 & 6,73 & 2,1 & 1,04 & 0,75 & 3.2*** & -0.00 & -0,09 & 830 & 5 & 10 & VIX9D & 21 & 11,52 & 3,12 & 1,39 & 0,08 & 3.69*** & -0,02 & -0,16 & 917 \\
3 & 10 & VIX9D & 42 & 5,28 & 1,27 & 1,01 & 0,75 & 4.14*** & -0.00 & -0,08 & 854 & 5 & 10 & VIX9D & 42 & 9,08 & 1,91 & 0,74 & 0,12 & 4.74*** & -0,03 & -0,14 & 940 \\
3 & 10 & VIX9D & 63 & 4,87 & 0,94 & 0,67 & 0,48 & 5.21*** & -0.00 & -0,06 & 865 & 5 & 10 & VIX9D & 63 & 8,12 & 1,51 & 0,93 & 0,22 & 5.38*** & -0,03 & -0,14 & 953 \\
3 & 10 & VIX9D & 84 & 4,66 & 0,77 & 0,5 & 0,37 & 6.07*** & -0.00 & -0,05 & 870 & 5 & 10 & VIX9D & 84 & 7,59 & 1,3 & 0,86 & 0,21 & 5.82*** & -0,03 & -0,13 & 956 \\
3 & 10 & VIX9D & 126 & 4,33 & 0,58 & 0,36 & 0,25 & 7.44*** & -0.00 & -0,04 & 871 & 5 & 10 & VIX9D & 126 & 6,94 & 1,03 & 0,58 & 0,14 & 6.73*** & -0,03 & -0,11 & 958 \\
3 & 10 & VIX9D & 252 & 3,83 & 0,47 & 0,37 & 0,37 & 8.14*** & -0.00 & -0,04 & 878 & 5 & 10 & VIX9D & 252 & 5,97 & 0,79 & 0,39 & 0,15 & 7.54*** & -0,02 & -0,09 & 963 \\ \arrayrulecolor{black}\bottomrule
\end{tabular}%
}
\source{Column N contains the total number of positions opened by the specified strategy. Moneyness (\%OTM) is expressed as a percentage of out-of-the-money. *, **, and *** denote statistical significance at the 10\%, 5\%, and 1\% levels, respectively, from the probabilistic Sharpe Ratio test against the benchmark buy-and-hold (B\&H) strategy. The VIX memory values are expressed in number of days used to calculate the percentile rank of the most recent VIX value.}
\end{table}

\clearpage

\section{Short Put Strategies with Kelly-VIX Combination Sizing}
\begin{table}[!ht]
\caption{Results of Short Put Strategies with Kelly-VIX Combination Sizing with VIX30D}
\label{tab:kelly-vix-30d-results}
\scriptsize
\resizebox{\columnwidth}{!}{%
\begin{tabular}{ccccccccccccc|ccccccccccccc}
\hline
\textbf{DTE} &
  \textbf{\%OTM} &
  \makecell{\textbf{VIX}\\\textbf{MEM.}} &
  \textbf{EST.} &
  \makecell{\textbf{EST.}\\\textbf{MEM.}} &
  \makecell{\textbf{aRC}\\\textbf{(\%)}} &
  \makecell{\textbf{aSD}\\\textbf{(\%)}} &
  \makecell{\textbf{MD}\\\textbf{(\%)}} &
  \textbf{MLD} &
  \makecell{\textbf{IR}\\\textbf{(\%)}} &
  \makecell{\textbf{VaR}\\\textbf{(\%)}} &
  \makecell{\textbf{CVaR}\\\textbf{(\%)}} &
  \textbf{N} &
  \textbf{DTE} &
  \textbf{\%OTM} &
  \makecell{\textbf{VIX}\\\textbf{MEM.}} &
  \textbf{EST.} &
  \makecell{\textbf{EST.}\\\textbf{MEM.}} &
  \makecell{\textbf{aRC}\\\textbf{(\%)}} &
  \makecell{\textbf{aSD}\\\textbf{(\%)}} &
  \makecell{\textbf{MD}\\\textbf{(\%)}} &
  \textbf{MLD} &
  \textbf{IR} &
  \makecell{\textbf{VaR}\\\textbf{(\%)}} &
  \makecell{\textbf{CVaR}\\\textbf{(\%)}} &
  \textbf{N} \\ \hline
0            & 0            & 21                  & GK                 & 5                         & 1,72         & 12,23        & 24,37       & 1,36         & 0.14        & -0,12        & -1,74         & 218        & 3            & 0            & 21                  & GK                 & 5                         & 23,52        & 31,1         & 35,43       & 1,17         & 0.76        & -2,81        & -5,82         & 628        \\
0            & 0            & 21                  & GK                 & 63                        & 1,35         & 10,63        & 22,69       & 1,36         & 0.13        & -0,02        & -1,26         & 169        & 3            & 0            & 21                  & GK                 & 63                        & 12,47        & 20,09        & 18,62       & 0,8          & 0.62        & -1,24        & -3,47         & 450        \\
\hline
\end{tabular}%
}
\source{This table presents results of strategies with Kelly-VIX30D combination sizing. Column N contains the total number of positions opened by the specified strategy. Moneyness (\%OTM) is expressed as a percentage of out-of-the-money. *, **, and *** denote statistical significance at the 10\%, 5\%, and 1\% levels, respectively, from the probabilistic Sharpe Ratio test against the benchmark buy-and-hold (B\&H) strategy. The VIX and estimator memory values are expressed in number of days used to calculate the percentile rank of the most recent VIX value.}
\end{table}
\begin{table}[!ht]
\caption{Results of Short Put Strategies with Kelly-VIX Combination Sizing with VIX9D}
\label{tab:kelly-vix-9d-results}
\small
\resizebox{\columnwidth}{!}{%
\begin{tabular}{ccccccccccccc|ccccccccccccc}
\hline
\textbf{DTE} &
  \textbf{\%OTM} &
  \makecell{\textbf{VIX}\\\textbf{MEM.}} &
  \textbf{EST.} &
  \makecell{\textbf{EST.}\\\textbf{MEM.}} &
  \makecell{\textbf{aRC}\\\textbf{(\%)}} &
  \makecell{\textbf{aSD}\\\textbf{(\%)}} &
  \makecell{\textbf{MD}\\\textbf{(\%)}} &
  \textbf{MLD} &
  \makecell{\textbf{IR}\\\textbf{(\%)}} &
  \makecell{\textbf{VaR}\\\textbf{(\%)}} &
  \makecell{\textbf{CVaR}\\\textbf{(\%)}} &
  \textbf{N} &
  \textbf{DTE} &
  \textbf{\%OTM} &
  \makecell{\textbf{VIX}\\\textbf{MEM.}} &
  \textbf{EST.} &
  \makecell{\textbf{EST.}\\\textbf{MEM.}} &
  \makecell{\textbf{aRC}\\\textbf{(\%)}} &
  \makecell{\textbf{aSD}\\\textbf{(\%)}} &
  \makecell{\textbf{MD}\\\textbf{(\%)}} &
  \textbf{MLD} &
  \textbf{IR} &
  \makecell{\textbf{VaR}\\\textbf{(\%)}} &
  \makecell{\textbf{CVaR}\\\textbf{(\%)}} &
  \textbf{N} \\ \hline
0            & 0            & 21                  & GK                 & 5                         & 9,34         & 9,95         & 9,47        & 0,49         & 0.94        & -0,09        & -1,19         & 220        & 3            & 0            & 21                  & GK                 & 5                         & 30,05        & 28,02        & 35,45       & 1,18         & 1.07        & -2,17        & -5,08         & 640        \\
0            & 0            & 21                  & GK                 & 63                        & 6,44         & 9,22         & 10,27       & 0,5          & 0.7         & -0,02        & -0,98         & 168        & 3            & 0            & 21                  & GK                 & 63                        & 15,63        & 18,87        & 17,66       & 0,58         & 0.83        & -1,02        & -3,13         & 466        \\
\hline
\end{tabular}%
}
\source{This table presents results of strategies with Kelly-VIX9D combination sizing. Column N contains the total number of positions opened by the specified strategy. Moneyness (\%OTM) is expressed as a percentage of out-of-the-money. *, **, and *** denote statistical significance at the 10\%, 5\%, and 1\% levels, respectively, from the probabilistic Sharpe Ratio test against the benchmark buy-and-hold (B\&H) strategy. The VIX and estimator memory values are expressed in number of days used to calculate the percentile rank of the most recent VIX value.}
\end{table}

\end{document}